\documentclass[prd,nofootinbib,twocolumn]{revtex4}

\usepackage{color}
\usepackage[squaren]{SIunits}
\usepackage{amsmath}
\usepackage{amsfonts}
\usepackage{slashed}
\usepackage{graphicx}
\usepackage{subfigure}
\usepackage{rotating}

\usepackage{graphicx,amsmath,amssymb,epsfig,verbatim,mathrsfs,array,layout,textcomp,latexsym}
\usepackage{multirow}
\usepackage{ulem}

\allowdisplaybreaks[1]

\renewcommand{\(}{\left(}
\renewcommand{\)}{\right)}
\renewcommand{\[}{\left\lbrack}
\renewcommand{\]}{\right\rbrack}
\renewcommand{\Re}[1]{\mathrm{Re}\left(#1\right)}
\renewcommand{\Im}[1]{\mathrm{Im}\left(#1\right)}

\newcommand{\nn}{\nonumber}

\newcommand{\cf}{cf.\,}
\newcommand{\refeq}[1]{Eq.~(\ref{eq:#1})}
\newcommand{\refeqs}[2]{Eqs.~(\ref{eq:#1})-(\ref{eq:#2})}
\newcommand{\reffig}[1]{Fig.~\ref{fig:#1}}
\newcommand{\refsec}[1]{Section \ref{sec:#1}}
\newcommand{\reftab}[1]{Table~\ref{tab:#1}}
\newcommand{\refapp}[1]{Appendix~\ref{sec:#1}}
\newcommand{\order}[1]{\mathcal{O}\({#1}\)}
\newcommand{\para}{\parallel}

\def \re{\textrm{Re}}
\def \im{\textrm{Im}}

\newcommand{\GeV}{\,\mathrm{GeV}}
\newcommand{\MeV}{\,\mathrm{MeV}}

\newcommand{\wilson}[2][{}]{\mathcal{C}_{#2}^{\mathrm{#1}}}

\newcommand{\braket}[1]{\left\langle #1 \right\rangle}

\def \Op{{\cal O}}

\def \alE{\alpha_e}        
\def \GF{G_F}              

\def \azeL{{A_0^L}}
\def \azeR{{A_0^R}}
\def \apaL{{A_\|^L}}
\def \apaR{{A_\|^R}}
\def \apeL{{A_\bot^L}}
\def \apeR{{A_\bot^R}}

\def \thl {{\theta_\ell}}
\def \thK {{\theta_{K}}}

\newcommand\tabvsptop{\rule{0pt}{2.6ex}}
\newcommand\tabvspbot{\rule[-1.5ex]{0pt}{0pt}}

\begin{document}

\setlength{\parindent}{0pt}

\begin{flushright}
DO-TH 12/16\\
EOS-2012-02\\
SI-HEP-2012-21\\
QFET-2012-02
\end{flushright}

\vspace*{-60mm}

\title{\boldmath General Analysis of 
       $\bar{B}\to\bar{K}^{(*)} \ell^+\ell^-$ Decays at Low Recoil}

\author{Christoph Bobeth} \affiliation{
       Excellence Cluster Universe, Technische Universit\"at M\"unchen,
       D-85748 Garching, Germany}
\author{Gudrun Hiller} \affiliation{Institut f\"ur Physik, Technische
       Universit\"at Dortmund, D-44221 Dortmund, Germany}
\author{Danny van Dyk} \affiliation{Institut f\"ur Physik, Technische
       Universit\"at Dortmund, D-44221 Dortmund, Germany}
       \affiliation{Theoretische Physik 1, Naturwissenschaftlich-Technische Fakult\"at,
       Universit\"at Siegen, Walter-Flex-Stra\ss{}e 3, D-57068 Siegen, Germany}

\begin{abstract}
  We analyze the angular distributions of $\bar{B}\to\bar{K}^* (\to \bar{K}\pi)
  \ell^+\ell^-$ and $\bar{B}\to\bar{K} \ell^+\ell^-$ decays in the region of low
  hadronic recoil in a model-independent way by taking into account the complete
  set of dimension-six operators $[\bar{s} \Gamma b][\bar{\ell} \Gamma' \ell]$.
  We obtain several novel low-recoil observables with high sensitivity to
  non-standard-model Dirac structures, including CP-asymmetries which do not
  require flavor tagging. The transversity observables $H_T^{(1,3,4,5)}$ are found to be
  insensitive to hadronic matrix elements and their uncertainties even when
  considering the complete set of operators.  In the most general scenario we show that
  the low recoil operator product expansion can be probed at the few-percent
  level using the angular observable $J_7$. Higher sensitivities are
  possible assuming no tensor contributions, specifically by testing the low-recoil
  relation $|H_T^{(1)}|=1$. We explicitly demonstrate the gain in reach of
  the low-recoil observables in accessing the ratio $|\wilson{9}/\wilson{10}|$
  compared to the forward-backward asymmetry, and probing CP-violating right-handed
  currents $\im\, \wilson{10\prime}$. We give updated Standard Model predictions for key
  observables in $\bar{B}\to\bar{K}^{(*)}\ell^+\ell^-$ decays.
\end{abstract} 

\maketitle

%
%
\section{Introduction}

Flavor Changing Neutral Current (FCNC) decays of beauty hadrons have a high
sensitivity to New Physics (NP) since the corresponding Standard Model (SM)
contributions are loop and flavor suppressed. In addition, the large value of
the $b$-quark mass facilitates the control of power corrections.

The large number of complementary observables and the excellent accessibility at
contemporary high energy experiments, in particular for muons, highlights the
exclusive FCNC decays $\bar{B}\to\bar{K}^*(\to \bar{K}\pi)\ell^+\ell^-$. In the
kinematic region of low hadronic recoil, where the emitted $K^*$ is soft in the
$B$-rest frame, a local Operator Product Expansion (OPE) can be performed
\cite{Grinstein:2004vb, Beylich:2011aq}. Together with the improved Isgur-Wise
relations \cite{Grinstein:2002cz, Grinstein:2004vb,Bobeth:2010wg}, this results
in a simple structure of the transversity decay amplitudes at leading order in
$1/m_b$ \cite{Bobeth:2010wg}
\begin{align} 
  \label{eq:universal}
  A_{i}^{L,R} &\propto  C^{L,R} f_i, & i &= \perp,\, ||,\, 0\, ,
\end{align}
factorizing into universal short-distance coefficients $C^{L,R}$ and form
factors $f_i$. This feature allows to extract short-distance couplings without
long-distance pollution, and vice versa, as well as to test the performance of
the OPE \cite{Bobeth:2010wg, Bobeth:2011gi}.

The opposite kinematical region of large recoil has been subjected to the
question of optimized observables as well, {\it e.g.,} \cite{Kruger:2005ep,
  Bobeth:2008ij, Egede:2008uy, Altmannshofer:2008dz, Lunghi:2010tr,
  Becirevic:2011bp, Matias:2012xw, Das:2012kz}. Several proposals exploit
specifically that QCD factorization (QCDF) \cite{Beneke:2001at, Beneke:2004dp} at
leading order maintains universal short-distance coefficients for
$A_{\perp}^{L,R}$ and $A_{||}^{L,R} $, while Eq.~(\ref{eq:universal}) is broken
for $i=0$ at lowest order, and for all $i=\perp,\,||,\,0$ at order $1/m_b$.
   
The additional benefit of the low recoil region is the strong parametric
suppression of the subleading $1/m_b$ corrections to the decay amplitudes at the
order of a few percent \cite{Grinstein:2004vb,Bobeth:2010wg}. Together with
an angular analysis \cite{Kruger:1999xa} this enables a rich flavor physics
program, complementing the large recoil region. One application is to extract
form factor ratios $f_i/f_j$ from data, as has recently been demonstrated in
\cite{Hambrock:2012dg, Beaujean:2012uj}.

The key questions addressed in this work are:
\begin{enumerate}
\item[\it i)] To which extent is Eq.~(\ref{eq:universal}) and its benefits
  preserved in the presence of operators beyond the SM ones?
\item[\it ii)] What are the optimal low recoil observables model-independently?
\item[\it iii)] What is their sensitivity to NP?
\item[\it iv)] What is the sensitivity to potential corrections to the OPE?
\end{enumerate}

To answer the above questions we perform a most general, model-independent
analysis of the decays $\bar{B}\to\bar{K}^*(\to \bar{K} \pi)\ell^+\ell^-$ and
$\bar{B}\to \bar{K}\ell^+\ell^-$. In terms of semileptonic dimension-six
operators $\[\bar{s}\,\Gamma\, b\]\[\bar\ell\, \Gamma^\prime\,\ell\]$ this
concerns the chirality-flipped partners of the SM ones, (pseudo-)scalar and
tensor operators. We compute various decay distributions and asymmetries.

The plan of the paper is as follows: The effective theory including the operator
basis is given in Section \ref{sec:eff:Ham}. We present low recoil observables
and relations from different operator sets in Section \ref{sec:lowreco:pheno}
and Section \ref{sec:BKll} for $\bar{B}\to\bar{K}^*(\to \bar{K} \pi)
\ell^+\ell^-$ and $\bar{B}\to \bar{K}\ell^+\ell^-$, respectively.  In Section
\ref{sec:sensitivity} we study the sensitivity of the low recoil observables to
even small NP effects. The sensitivity to OPE corrections is worked out in
Section \ref{sec:opebreak} as well as
  a brief discussion of S-wave backgrounds.  We conclude in Section \ref{sec:conclusion}.

In several appendices we give formulae and subsidiary information.  In
\refapp{ang:dist} we discuss the full angular decay distribution in
$\bar{B}\to\bar{K}^* (\to \bar{K}\pi) \ell^+\ell^-$ decays.  In \refapp{ang:obs}
we present the angular observables in terms of the transversity amplitudes for
the complete set of semileptonic $|\Delta B| = |\Delta S| = 1$ operators. In
\refapp{calculation} we detail the transversity amplitudes that parametrize the
tensor contribution to the matrix element.  An update of the SM predictions for
the key observables in $\bar{B}\to\bar{K}^*\ell^+\ell^-$ and
$\bar{B}\to\bar{K}\ell^+\ell^-$ decays is given in Appendix~\ref{sec:update}.

%
%
\section{The Effective Hamiltonian \label{sec:eff:Ham}}

Rare semileptonic $|\Delta B| = |\Delta S| = 1$ decays
are described by an effective Hamiltonian
\begin{align}
  \label{eq:Heff}
  {\cal{H}}_{\rm eff}= 
   - \frac{4\, G_F}{\sqrt{2}}  V_{tb}^{} V_{ts}^\ast \,\frac{\alE}{4 \pi}\,
     \sum_i \wilson[]{i}(\mu)  \Op_i(\mu).
\end{align}
Here, $G_F$ denotes Fermi's constant, $\alE$ the fine structure constant and
unitarity of the Cabibbo-Kobayashi-Maskawa (CKM) matrix $V$ has been used. The
subleading contribution proportional to $V_{ub}^{} V_{us}^\ast$ has been
neglected.

The renormalization scale $\mu$, which appears in the short-distance couplings
$\wilson[]{i}$ and the matrix elements of the operators $\Op_i$, is of the order
of the $b$-quark mass. In the following we suppress the dependence of the Wilson
coefficients $\wilson{i}$ on the scale $\mu$.

In the SM $b \to s\,\ell^+\ell^-$ processes are mainly governed by the operators
$\Op_{7,9,10}$ which will be referred to as the SM operator basis. Beyond the SM
chirality-flipped ones $\Op_{7',9',10'}$, collectively denoted here by SM', may
appear. The SM and SM' operators are written as \cite{Bobeth:2007dw,
  Altmannshofer:2008dz, Kruger:2005ep}
\begin{equation}
\begin{aligned}
  \Op_{7(7')} & = \frac{m_b}{e}\!\[\bar{s} \sigma^{\mu\nu} P_{R(L)} b\] F_{\mu\nu}\,,
\\
  \Op_{9(9')} & = \[\bar{s} \gamma_\mu P_{L(R)} b\]\!\[\bar{\ell} \gamma^\mu \ell\]\,,
\\[0.1cm]
  \Op_{10(10')} & = \[\bar{s} \gamma_\mu P_{L(R)} b\]\!\[\bar{\ell} \gamma^\mu \gamma_5 \ell\]\,.
\end{aligned}
\label{eq:SM:ops}
\end{equation}
Furthermore, we allow for scalar and pseudo-scalar operators, referred to as S
and P,
\begin{equation}
\begin{aligned}
    \Op_{S(S')}   & = \[\bar{s} P_{R(L)} b\]\!\[\bar{\ell} \ell\]\,,
\\[0.1cm]
    \Op_{P(P')}   & = \[\bar{s} P_{R(L)} b\]\!\[\bar{\ell} \gamma_5 \ell\]\,,
\end{aligned}
\label{eq:psd-scalar:ops}
\end{equation}
which includes the chirality-flipped ones, as well as tensor operators, referred
to as T and T5,
\begin{equation}
\begin{aligned}
  \Op_T   & = \[\bar{s} \sigma_{\mu\nu} b\]\!\[\bar{\ell} \sigma^{\mu\nu} \ell\]\,,
\\
  \Op_{T5} & = \[\bar{s} \sigma_{\mu\nu} b\] \[\bar{\ell} \sigma^{\mu\nu} \gamma_5 \ell\]\,.
\end{aligned}
\label{eq:tensor:ops}
\end{equation}
Note that $\Op_{T5} = - i/2\, \varepsilon^{\mu\nu\alpha\beta}
    \[\bar{s} \sigma_{\mu\nu} b\]\!\[\bar{\ell} \sigma_{\alpha\beta} \ell\]
  = - \Op_{TE}/2$,
see Eq.~(\ref{eq:gamma5rel}),
as commonly used in the literature \cite{Bobeth:2007dw,Kim:2007fx, Alok:2010zd}.

Current-current and QCD penguin operators $\Op_{i\leq 6}$, as well as the
chromo-magnetic dipole operator $\Op_{8}$ have to be included for a consistent
description of $b \to s \ell^+\ell^-$ decays, for definition see
\cite{Chetyrkin:1996vx}.  The matrix elements of $\Op_{1\dots6,8}$ contribute to
$b \to s + \{\gamma, g, \ell^+\ell^-\}$ processes via quark-loop effects. The
latter are taken into account by means of the effective Wilson coefficients
$\wilson[eff]{7,8,9}$. The effective Wilson coefficients are renormalization
group invariant up to higher orders in the strong coupling constant $\alpha_s$. In the case
of exclusive decays the $1/m_b$ corrections in the large- and low-recoil region
from QCDF  \cite{Beneke:2001at, Beneke:2004dp, Bobeth:2007dw}
or SCET \cite{Ali:2006ew, Lee:2006gs} and the low-recoil OPE
\cite{Grinstein:2004vb, Bobeth:2010wg, Bobeth:2011gi}, respectively, should be
included in the $\wilson[eff]{i}$.
We evaluate $\alpha_e$ at $\mu = \mu_b = \order{m_b}$ which takes into account
most of the NLO QED corrections \cite{Bobeth:2003at, Huber:2005ig}.

%
%
\section{$\bar{B}\to\bar{K}^{*} \ell^+\ell^-$ at low recoil \label{sec:lowreco:pheno}}

We study $\bar{B}\to\bar{K}^*(\to \bar{K} \pi)\ell^+\ell^-$ decays in the low
recoil region for a generalized operator basis and detail the relevant
observables and their relations. In Section \ref{sec:sm:ops} we give the results
using SM operators only. In Section \ref{sec:chi-flip:ops},
\ref{sec:scalar:ops}, \ref{sec:tensor:ops} we include either SM', S and P
or T and T5 operators, respectively. Interference effects are worked
out in Section \ref{sec:interference}.

The main results of this section are summarized in \reftab{overview}, where the
low recoil relations between the observables and the amount of their violations
is given. Our results are based on the angular distribution presented in
\refapp{ang:dist}, and the angular observables in \refapp{ang:obs}. 

%
\subsection{SM operators \label{sec:sm:ops}}

The amplitude of the exclusive decays $\bar{B} \to \bar{K}^* \ell^+\ell^-$ can
be treated at low recoil using an OPE and further matching onto HQET
\cite{Grinstein:2004vb}. After application of the improved Isgur-Wise relations
\cite{Grinstein:2004vb}, one finds for the transversity amplitudes
\cite{Bobeth:2010wg, Bobeth:2011gi}, see also Eq.~(\ref{eq:universal}),
\begin{align}
  \label{eq:Aloreco}
  A^{L,R}_{0,\parallel} & = -C^{L,R}\, f_{0,\parallel} \, , &
  A^{L,R}_{\perp}       & = +C^{L,R}\, f_{\perp} \, .
\end{align} 
The short-distance coefficients read
\begin{align}
  C^{L,R}(q^2) & 
    = \wilson[eff]{79}(q^2) \mp \wilson[]{10},  
\\[0.1cm]
  \wilson[eff]{79}(q^2) &
    = \wilson[]{9} + \kappa \frac{2\, m_b M_B}{q^2} \wilson[]{7} + Y(q^2),
    \label{eq:C79}
\end{align}
where $Y$ denotes the matrix elements of the 4-quark operators, see
\cite{Bobeth:2011gi} for details. Here, the matching correction $\kappa=1-2
\alpha_s/(3 \pi) \ln \mu/m_b +{\cal{O}}(\alpha_s^2)$ arises from the lowest
order improved Isgur-Wise relations. Its $\mu$-dependence compensates the one of
the dipole form factors $T_{1,2,3}$.

The term $\propto \wilson[]{7}$ in \refeq{C79} involves uncertainties from
corrections at order $1/m_b$. However, since generically $|\wilson[]{9,10}| \gg
|\wilson[]{7}|$ (in the SM $\wilson[]{9}=4.2$, $\wilson[]{10}=-4.2$ and
$\wilson[]{7}= -0.3$) the coefficient $C^{L}$ can be regarded as strongly
short-distance dominated whereas $C^{R}$ yields only a numerically subleading
contribution to observables. It follows that the subleading power corrections
enter the amplitude at the few percent level.

The form factors $f_i$, also termed helicity form factors
\cite{Bharucha:2010im}, can be written in terms of the usual heavy-to-light
vector and axial-vector form factors $V$, $A_{1,2}$ as \cite{Bobeth:2010wg}
\begin{equation}
\begin{aligned}
  \frac{f_{\perp}}{N} & = \frac{\sqrt{2\, \lambda}}{M_B + M_{K^*}} V\,, 
\\[0.2cm]
  \frac{f_{\parallel}}{N} & = \sqrt{2}\, (M_B + M_{K^*})\, A_1\,,
\\[0.2cm]
  \frac{f_{0}}{N} & = \frac{(M_B^2 - M_{K^*}^2 - q^2) (M_B + M_{K^*})^2 A_1
   - \lambda\, A_2}{2\, M_{K^*} (M_B + M_{K^*}) \sqrt{q^2}}\,.
\end{aligned}
\label{eq:f0-pp-pa:FFs}
\end{equation}
The normalization factor $N$ depends on the invariant mass squared of the lepton
pair, $q^2$, and is given in \refeq{trAmp:norm:factor}. The kinematical factor
$\lambda \equiv \lambda(M_B^2, M_{K^*}^2, q^2)$ is given in \refeq{defLambda}.

The factorization into short-distance coefficients and form factors,
\refeq{Aloreco}, allows to identify suitable combinations of the observables
$J_i$ appearing in the angular distribution of $\bar{B}\to \bar{K}^*(\to \bar{K}
\pi) \ell^+\ell^-$, see \refapp{ang:dist} for details.  The angular observables
depend on two short-distance parameters $\rho_{1,2}$ only,
\begin{equation}
\begin{aligned}
  \frac{4}{3 \beta_\ell^2} (2 J_{2s} &+ J_3) = 2\, \rho_{1} f_\perp^2, &
  - \frac{4}{3 \beta_\ell^2} J_{2c} & = 2\, \rho_{1} f_0^2\,,
\\[0.1cm]
  \frac{4}{3 \beta_\ell^2} (2 J_{2s}&  - J_3)  = 2\, \rho_{1} f_\parallel^2, &
  \frac{4 \sqrt{2}}{3 \beta_\ell^2} J_4 & = 2\, \rho_{1} f_0 f_\parallel\,,
\\[0.1cm]
    \frac{2 \sqrt{2}}{3 \beta_\ell} J_5 & = 4\, \rho_2 f_0 f_\perp, & 
      \frac{2}{3 \beta_\ell} J_{6s} & = 4\, \rho_2 f_\parallel f_\perp\,,
\end{aligned}
\label{eq:hiQ2:J:SM}
\end{equation}
where
\begin{align}
  \rho_1 & 
  = \frac{1}{2} \left(|C^R|^2 + |C^L|^2 \right) 
  = \Big| \wilson[eff]{79} \Big|^2 + \Big| \wilson[]{10} \Big|^2,
\\[0.2cm]
  \rho_2 & 
  = \frac{1}{4} \left(|C^R|^2 - |C^L|^2 \right)
  = \mbox{Re} \left( \wilson[eff]{79}\, \wilson[*]{10} \right) .
\end{align}
Note that $J_{7,8,9} = 0$ \cite{Bobeth:2010wg} and $J_{6c}=0$ since neither S, P
\cite{Altmannshofer:2008dz} nor T, T5 operators are present.

From \refeq{hiQ2:J:SM} follow \cite{Bobeth:2010wg} the short- and long-distance
free ratio
\begin{align}
  H_T^{(1)} & 
  \equiv \frac{\sqrt{2} J_4}{\sqrt{-J_{2c} \left(2 J_{2s} - J_3\right)}},
\end{align}
as well as the long-distance free ratios
\begin{align}
  \label{eq:def:HT2:J2c}
  H_T^{(2)} &
  \equiv \frac{\beta_\ell J_5}{\sqrt{-2 J_{2c} \left(2 J_{2s} + J_3\right)}}, 
\\[0.1cm]  \label{eq:def:HT3:J2s}
  H_T^{(3)} & 
  \equiv \frac{\beta_\ell J_{6s}}{2 \sqrt{(2 J_{2s})^2 - J_3^2}}.
\end{align}
Here we point out a further nontrivial observable, which does depend
neither on form factors nor on short-distance physics:
\begin{align} \label{eq:HT1b}
  H_T^{(1b)} & 
  \equiv -\frac{J_{2c}\, J_{6s}}{2\,J_4\, J_5} ,
\end{align}
and which equals  one.
Note that this observable can be obtained via $H_T^{(1b)}=H_T^{(3)}/[H_T^{(1)} H_T^{(2)}]$.
However, by using the definition Eq.~(\ref{eq:HT1b}) directly  different
$J_i$ appear. This offers additional advantages in the experimental extraction
from the angular distributions.

In addition, long-distance free CP asymmetries $a_{\rm CP}^{(1,2,3)}$ can be
formed, which are related to the CP asymmetry of the decay rate, of the
forward-backward asymmetry, and of $H_T^{(2,3)}$, respectively
\cite{Bobeth:2011gi}.

Furthermore, several short-distance free ratios of form factors
(\ref{eq:f0-pp-pa:FFs}) can be obtained
\begin{align}
  \label{eq:sd-free:ffFirst}
  \frac{f_0}{f_\parallel}
  & = \frac{\sqrt{2} J_5}{J_{6s}}
    = \frac{-J_{2c}}{\sqrt{2} J_4}
\\ \label{eq:sd-free:ffMid}
  & = \frac{\sqrt{2} J_4}{2 J_{2s} - J_3}
    = \sqrt{\frac{-J_{2c}}{2 J_{2s} - J_3}}\,,
\nonumber\\[0.1cm]
  \frac{f_\perp}{f_\parallel}
  & = \sqrt{\frac{2 J_{2s} + J_3}{2 J_{2s} - J_3}}
    = \frac{\sqrt{-J_{2c} \left(2 J_{2s} + J_3\right)}}{\sqrt{2} J_4}\,,
\\[0.1cm]
  \label{eq:sd-free:ffLast}
  \frac{f_0}{f_\perp}
  & = \sqrt{\frac{-J_{2c}}{2 J_{2s} + J_3}}\,.
\end{align}
They allow to extract information on form factors directly from the data
\cite{Hambrock:2012dg, Beaujean:2012uj}, providing a benchmark test for form
factor determinations such as from lattice QCD.

To sum up, using SM-type operators only -- which may or may not receive
contributions from beyond the SM -- the low recoil OPE predicts at leading order
in $1/m_b$
\begin{align}
  \frac{ H_T^{(1)} }{{\rm sgn}(f_0)} = H_T^{(1b)} & = 1\,, 
  & J_{7,8,9} & = 0\,,
\nonumber\\[0.1cm]
  H_T^{(2)} = H_T^{(3)} & = 2\, \frac{\rho_2}{\rho_1}\,,
\end{align}
and the observable form factor ratios given in
\refeqs{sd-free:ffFirst}{sd-free:ffLast}.  As already stressed the subleading
power corrections are parametrically suppressed and at the few percent level.

%
\subsection{Chirality-flipped operators \label{sec:chi-flip:ops}}

\begin{figure}[t]
\subfigure[\label{fig:ht1}]{\includegraphics[width=0.48\textwidth]{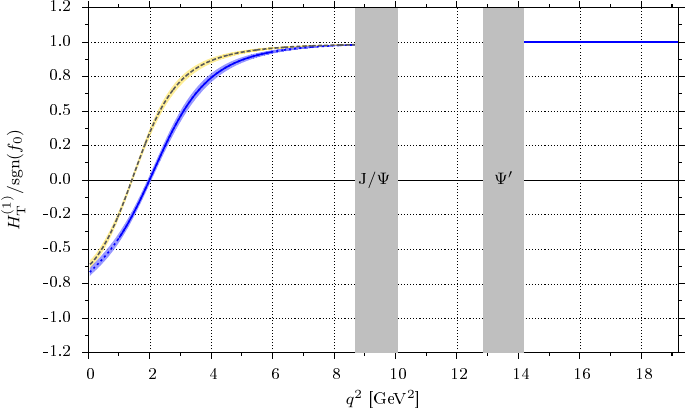}}\\
\subfigure[\label{fig:ht1b}]{\includegraphics[width=0.48\textwidth]{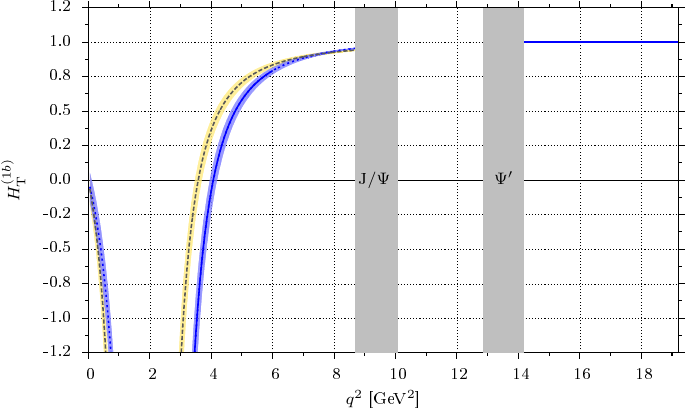}}
\caption{
  $H_{ T}^{(1)}/{\rm sgn}(f_0)$ (a) and $H_T^{(1b)}$ (b) 
   in the
 large and low recoil regions, below and above the experimentally
 vetoed narrow charmonium backgrounds (vertical grey bands)
 from $\bar{B}\to J/\Psi( \to \ell^+\ell^-)\bar{K}^*$ and
 $\bar{B}\to \Psi'( \to \ell^+\ell^-)\bar{K}^*$, respectively.
  Shown are the SM prediction (blue solid) and the (SM+SM') benchmark
 point (black dashed) with their respective
 uncertainty bands (darker (blue) and lighter (gold)),
 respectively. See text for details.
   \label{fig:ht1}
}
\end{figure}

Taking into account the chirality flipped operators the universal structure of
the transversity amplitudes (\ref{eq:Aloreco}) is broken in part. One obtains in
the (SM+SM') model
\begin{align}
  A^{L,R}_{0,\parallel} & = -C_{-}^{L,R}\, f_{0,\parallel}, & 
  A^{L,R}_\perp         & = +C_{+}^{L,R}\, f_\perp ,
\end{align}
where
\begin{align}
  \label{eq:Cminus}
  C_{-}^{L,R} (q^2)& = 
     \wilson[eff]{79} - \wilson[eff]{7'9'}  \mp (\wilson[]{10} - \wilson[]{10'}),
\\[0.1cm]
  \label{eq:Cplus}
  C_{+}^{L,R} (q^2) & =   \wilson[eff]{79} + \wilson[eff]{7'9'} 
    \mp (\wilson[]{10} + \wilson[]{10'}) , 
\\[0.2cm]
  \wilson[eff]{7'9'}(q^2) &
    = \wilson[]{9'} + \kappa \frac{2\, m_b M_B}{q^2} \wilson[]{7'} + Y'(q^2).
\end{align}
Here $\wilson[eff]{7'9'}$ is defined analogously to $\wilson[eff]{79}$, {\it
  i.e.}, $Y'$ denotes the matrix element of the chirality-flipped 4-quark
operators.

The angular observables $J_i$ in (SM+SM') read
\begin{align}  \nonumber  
  \frac{4}{3 \beta_\ell^2} (2 J_{2s} & + J_3)  = 2\, \rho_{1}^{+} f_\perp^2, &
   - \frac{4}{3 \beta_\ell^2} J_{2c} & = 2\, \rho_{1}^{-} f_0^2 ,\\   \label{eq:hiQ2:I}
     \frac{4}{3 \beta_\ell^2} (2 J_{2s} &  - J_3)  = 2\, \rho_{1}^{-} f_\parallel^2, &
  \frac{4 \sqrt{2}}{3 \beta_\ell^2} J_4 &  = 2\, \rho_{1}^{-} f_0 f_\parallel, 
\\[0.1cm] \nonumber
  \frac{2 \sqrt{2}}{3 \beta_\ell} J_5 & = 4\, {\rm Re}(\rho_2) f_0 f_\perp, &
 \frac{2}{3 \beta_\ell}  J_{6s} & = 4\, {\rm Re}(\rho_2) f_\parallel f_\perp,
\\[0.1cm]
  \frac{4 \sqrt{2}}{3 \beta_\ell^2} J_8 & = 4\, {\rm Im}(\rho_2) f_0 f_\perp, &
  - \frac{4}{3 \beta_\ell^2} J_9 & = 4\, {\rm Im}(\rho_2) f_\parallel f_\perp, &
  \nonumber
\end{align}
where $J_7=0$ still holds and $\rho_1$ and $\rho_2$ have been generalized to
\begin{equation}
\begin{aligned}
  \rho_1^{\pm} & \equiv \frac{1}{2} \left(|C^R_\pm|^2 + |C^L_\pm|^2 \right),
\\[0.1cm]
  \rho_2^{}    & \equiv \frac{1}{4} \left(C_{+}^R C_{-}^{R*}  - C_{-}^L C_{+}^{L*} \right).
\end{aligned}
\end{equation}
Switching off the chirality flipped operators one recovers $C_+^L = C_-^L =
C^L_{}$ (and analogously for $L \to R$), such that $\rho_1^+ = \rho_1^- =
\rho_1^{}$.

In (SM+SM'), the asymmetries $H_T^{(2,3)}$, defined in 
Eqs.~(\ref{eq:def:HT2:J2c})-(\ref{eq:def:HT3:J2s}), read
\begin{align}
  \label{eq:HT23}
  H_T^{(2)}
    & = 2\, \frac{{\rm Re}(\rho_2)}{\sqrt{\rho_1^- \cdot \rho_1^+}}\,, & 
  H_T^{(3)}
    & = 2\, \frac{{\rm Re}(\rho_2)}{\sqrt{\rho_1^- \cdot \rho_1^+}}\,.
\end{align} 
They remain long-distance free. Furthermore, the low recoil predictions obtained in the SM
basis
\begin{align}
 \frac{H_T^{(1)}}{\mbox{sgn}(f_0)} & = H_T^{(1b)} =1, & 
  H_T^{(2)} & = H_T^{(3)}, &
  J_7 & = 0
\end{align}
remain intact. 

In Fig.~\ref{fig:ht1} we show $H_T^{(1)}$ and $H_T^{(1b)}$. While both equal
one at low recoil in SM+SM', at large recoil both observables exhibit a
nontrivial $q^2$-dependence and depend on short- and long-distance contributions.
However, to lowest order  form factors drop out in $H_T^{(1)}$ {\it \cf}~\cite{Matias:2012xw}
and $H_T^{(1b)}$. We show the residual uncertainty from the form factors and
subleading $1/m_b$ corrections by the shaded bands. The SM is represented by
the thin (blue) band, whereas the lighter shaded (gold) one corresponds to a
scenario with $\wilson{7',9'}=\wilson[SM]{7,9}$, $\wilson{10'}=-\wilson[SM]{10}$,
$\wilson{7,9,10}=0$. For numerical input, see Appendix~\ref{sec:update}.

Since in (SM+SM') $J_{8,9} \neq 0$ two additional long-distance free ratios
\begin{align}
  \label{eq:HT45}  
  H_T^{(4)} &
  \equiv \frac{2  J_8}{\sqrt{-2 J_{2c} \left(2 J_{2s} + J_3 \right)}}\,,
\\[0.2cm]
  H_T^{(5)} &
  \equiv \frac{- J_9}{\sqrt{(2 J_{2s})^2 - J_3^2}} 
\end{align}
can be constructed. They obey
\begin{align}
  H_T^{(4)} & = H_T^{(5)}=2\, \frac{{\rm Im}(\rho_2)}{\sqrt{\rho_1^- \cdot \rho_1^+}}\,. 
\end{align}
We point out a further nontrivial observable, which depends neither on form
factors nor on short-distance physics:
\begin{align} \label{eq:HT1c}
  H_T^{(1c)} & 
  \equiv 2\,\frac{J_4\, J_8}{J_{2c}\, J_{9}} \,,
\end{align}
where in (SM+SM') 
\begin{align}
 H_T^{(1c)} & = 1 \,.
\end{align}
For $H_T^{(1c)}$ an analogous comment  as on $ H_T^{(1b)} $ applies, see text below
Eq.~(\ref{eq:HT1b}).

The transverse asymmetries $H_T^{(2,3,4,5)}$ are driven by the real and
imaginary part of $\rho_2$, written as
\begin{align}
  \mbox{Re}(\rho_2) & =  
  \mbox{Re} (\wilson[eff]{79} \, \wilson[*]{10} - \wilson[eff]{7'9'}\, \wilson[*]{10'})\,,
\\[0.1cm] \label{eq;imrho2}
  \mbox{Im}(\rho_2) & = 
  \mbox{Im} (\wilson[eff]{7'9'} \, \wilson[eff*]{79} - \wilson[]{10}\,\wilson[*]{10'})\,,
\end{align}
where $\mbox{Im}(\rho_2)$ vanishes for $\wilson[]{i'} = 0$ including vanishing
chirality-flipped four-quark operators. Real-valued SM+SM' Wilson coefficients can
still induce somewhat suppressed, finite values of $H_T^{(4,5)}$ through the
absorptive contributions in the matrix elements of the four-quark operators
$Y$ and $Y^\prime$, by $\mbox{Im}(\rho_2) = \mbox{Re}\, \wilson[eff]{79}
\,\mbox{Im}\, Y' - \mbox{Re}\,\wilson[eff]{7'9'} \,\mbox{Im}\, Y$. Note that in
the SM in the low recoil region $\mbox{Im}\, Y_{\rm SM} \sim 0.2 - 0.3$
\cite{Bobeth:2011gi}.  In any case, $H_T^{(4,5)}$ are null tests of the SM.
A SM background to right-handed currents arises at higher order in the OPE
including and counting $m_s/m_b$-terms as such and enters $H_T^{(4,5)}$
with additional parametric suppression by $\alpha_s$ or $\wilson{7}/\wilson{9}$
\cite{Grinstein:2004vb,Beylich:2011aq}.

Combinations result in further useful observables which do not depend
on form factors either:
\begin{align}
  \frac{H_T^{(4)}}{H_T^{(2)}} & =  \frac{2}{\beta_\ell} \frac{J_8}{J_5}\,, &
  \frac{H_T^{(5)}}{H_T^{(3)}} & = -\frac{2}{\beta_\ell} \frac{J_9}{J_{6s}}\,.
  \label{eq:J9J6}
\end{align}
Both equal ${\rm Im}(\rho_2)/{\rm Re}(\rho_2)$ in (SM+SM').

Since $J_{8,9}$ are naive T-odd these angular observables give optimal access to
CP violation in the presence of small strong phases \cite{Bobeth:2008ij}.  Since
both $J_{8,9}$ are also CP-odd, $H_T^{(4,5)}$ can be measured from $B$-meson
samples without tagging and give rise to a further long-distance-free CP
asymmetry defined as
\begin{equation}
  a_{\rm CP}^{(4)} = 
  \begin{cases}
    \displaystyle
    \frac{\sqrt{2}(J_8\!-\!\bar{J}_8)}{\sqrt{-(J_{2c}\!+\!\bar{J}_{2c}) 
      \big[2 (J_{2s}\!+\!\bar{J}_{2s})\!+\!(J_3\!+\!\bar{J}_3)\!\big]}}
\\[0.1cm]
    \displaystyle
    -\frac{J_9 - \bar{J}_9}{\sqrt{4 \big(J_{2s} + \bar{J}_{2s}\big)^2 - 
      \big(J_3 + \bar{J}_3\big)^2}}
  \end{cases}
\end{equation}
for $H_T^{(4)}$ and $H_T^{(5)}$, respectively.  Here, the barred quantities are
obtained by conjugating the weak phases.  In terms of the short-distance
coefficients $a_{\rm CP}^{(4)}$ reads
\begin{align}
  a_{\rm CP}^{(4)} & = 
   2\, \frac{{\rm Im}(\rho_2 - \bar{\rho}_2)}
   {\sqrt{(\rho_1^+ + \bar{\rho}_1^+) \cdot (\rho_1^- + \bar{\rho}_1^-)}}\,.
\end{align}
The generalization of $a_{\rm CP}^{(3)}$ \cite{Bobeth:2011gi} is given by
\begin{align}
  a_{\rm CP}^{(3)} & = 
   2\, \frac{{\rm Re}(\rho_2 - \bar{\rho}_2)}
   {\sqrt{(\rho_1^+ + \bar{\rho}_1^+) \cdot (\rho_1^- + \bar{\rho}_1^-)}}\,.
\end{align}

Due to the presence of $\rho_1^+$ and $\rho_1^-$, the generalization of the CP
asymmetries $a_{\rm CP}^{(1)}$ and $a_{\rm CP}^{(2)}$ leads to a doubling
\begin{align}
  \label{eq:aCP1plus}
  a_{\rm CP}^{(1,\pm)} & \equiv 
    \frac{\rho_1^\pm - \bar\rho_1^\pm}{\rho_1^\pm + \bar\rho_1^\pm}\,, &
  a_{\rm CP}^{(2,\pm)} & \equiv 
    \frac{\displaystyle \frac{\rho_2}{\rho_1^\pm} - \frac{\bar\rho_2}{\bar\rho_1^\pm}}
         {\displaystyle \frac{\rho_2}{\rho_1^\pm} + \frac{\bar\rho_2}{\bar\rho_1^\pm}}\,.
\end{align}
In this case the CP asymmetry of the decay rate can not be related to any of the
$a_{\rm CP}^{(1,\pm)}$ and is not long-distance free. However, from
(\ref{eq:hiQ2:I}) it is straightforward to read off strategies to relate the
$a_{\rm CP}^{(k,\pm)}$, $k=1,2$ to the $J_i$. In particular $a_{\rm CP}^{(1,-)}$
can be extracted from ratios involving $J_{2c},\, (2 J_{2s} - J_3),\, J_4$,
whereas $a_{\rm CP}^{(1,+)}$ requires $(2 J_{2s} + J_3)$. In analogy to
Eq. (2.37) of Ref.~\cite{Bobeth:2011gi}, the set $b$ has to be restricted to $b =
\{1,3,4\}$ for $a_{\rm CP}^{(2,-)}$ and to $b = 2$ for $a_{\rm CP}^{(2,+)}$.

In (SM+SM') short-distance free ratios of  angular observables $J_i$ exist for
$f_0/f_\parallel$ as given in \refeq{sd-free:ffFirst}, and additionally
\begin{align} 
  \label{eq:ff-ratio-test}
  \frac{f_0}{f_\parallel} &
    = \frac{\sqrt{2} J_8}{-J_9}\,.
\end{align}
Due to $(2 J_{2s} + J_3) \propto \rho_1^+$, however, no short-distance free
ratios can be formed which involve $f_\perp$. Hence, the observables $F_L$ and
$A_T^{(2,3)}$ are no longer short-distance free \cite{Bobeth:2010wg} either
\begin{align}
 & F_L 
    = \frac{\rho_1^- f_0^2}{\rho_1^- (f_0^2 + f_\parallel^2) + \rho_1^+ f_\perp^2}\,, &
\\[0.1cm]
 & A_T^{(2)} 
    = \frac{\rho_1^+ f_\perp^2 - \rho_1^- f_\parallel^2}
           {\rho_1^+ f_\perp^2 + \rho_1^- f_\parallel^2}\,, &
  A_T^{(3)} &
    = \sqrt{\frac{\rho_1^-}{\rho_1^+}}\, \frac{f_\parallel}{f_\perp} \,, &
\end{align}
 and
the method used in \cite{Hambrock:2012dg} to extract form factor ratios would
yield $\sqrt{\rho_1^-/\rho_1^+} (f_\parallel/f_\perp)$. With current data the
correction factor is within $0.7 \leq  \sqrt{\rho_1^-/\rho_1^+} \leq 1.4$ at
2 $\sigma$.

Furthermore, we obtain the relation in (SM+SM')
\begin{align}
  A_T^{(3)} & = \sqrt{\big(1 - A_T^{(2)}\big)\Big/\big(1 + A_T^{(2)}\big)}\,,
\end{align}
which can be checked experimentally.

%
\subsection{Scalar and pseudo-scalar operators \label{sec:scalar:ops}}

The (S+P) operators modify the angular observables $J_{1c,5,6c,7}$ only.  The
respective NP contributions are driven by $A_0 \Delta_{S,P}$, where $A_0$
denotes the $B \to K^*$ axial-vector form factor and $\Delta_{S,P} \equiv
\wilson[]{S,P} - \wilson[]{S',P'}$.  We find that $J_{1c}$ only receives generically
unsuppressed contributions,
\begin{equation}
\begin{aligned} 
  J_{1c}
    & = \frac{3}{2} \rho_1 f_0^2 +
3 N^2( |\Delta_S |^2 + |\Delta_P |^2 )\frac{\lambda}{m_b^2} A_0^2 \\
&    + {\cal{O}}(m_{\ell}^2/q^2, m_s/m_b) \, .
\end{aligned}
\end{equation}
Helicity-suppressed ($\sim m_\ell/\sqrt{q^2}$) contributions from interference
terms SM$\,\times\,$S arise in $J_{5,6c,7}$.  For the explicit expressions
see \refapp{ang:obs}.

We find that in the presence of (S+P) operators the low recoil relations
\begin{equation}
\begin{aligned}
    |H_T^{(1)}| & = 1\,, & H_T^{(1b)} & = 1 + \order{\frac{m_\ell}{\sqrt{q^2}}}\,,\\
      H_T^{(4)} & = H_T^{(5)}
\end{aligned}
\end{equation}
hold, and $H_T^{(3,4,5)}$ remain long-distance free.  Since $J_{8,9}$ vanish in
the considered scenario $H_T^{(4)} = H_T^{(5)}=0$, and $H_T^{(1c)}$ is
ill-defined as in the SM-like scenario.

The helicity-suppressed contributions to $J_5$ break the relation $H_T^{(2)} =
H_T^{(3)}$ at ${\cal{O}}(m_\ell/\sqrt{q^2})$ through a finite $\Delta_S$. In
this case $H_T^{(2)}$ ceases to be free of form factors, and rather depends on
$A_0 / f_0$.  Moreover, the relation $J_7 = 0$ is broken at
${\cal{O}}(m_\ell/\sqrt{q^2})$ if there is additionally CP violation beyond the
SM. With the exception of using $J_5$, the ratio $f_0/f_\parallel$ can be
extracted by means of the methods proposed in Eqs.~(\ref{eq:sd-free:ffFirst})
and (\ref{eq:ff-ratio-test}).

The (pseudo-)scalar contributions to $J_{1c}$ break the relation $J_{1c} =
-J_{2c}$, valid only in the (SM+SM') basis for $m_\ell \to 0$, see also
\refapp{ang:obs}. At the same time, contributions to the longitudinal
polarization $F_{L}$ of $K^*$ mesons are induced, see \refeq{defFL}. These
contributions prohibit that $F_L$ and $A_{\rm FB}$, the lepton forward-backward
asymmetry, can be extracted simultaneously from a fit to \refeq{dGdcL}, the
angular distribution in $\cos \thl$. Note that $F_L$ and $F_T=1-F_L$ can be
extracted from \refeq{dGdcK}, the distribution in $\cos \thK$, in a
model-independent way. Discrepancies between the extracted values of $F_{L,T}$
from Eqs.~(\ref{eq:dGdcK}) and (\ref{eq:dGdcL})
would indicate BSM physics.  (We assume here that S-wave contributions from
$\bar B \to \bar K \pi \ell^+ \ell^-$ have been removed from the data, see 
Section \ref{sec:Swave}.)  Note also that interference terms
(SM+SM')$\,\times\,$S contribute to $A_{\rm FB}$ via $J_{6c}$ due to (\ref{eq:afb}).

%
\subsection{Tensor operators \label{sec:tensor:ops}}

The  tensor operators (T+T5) give rise to additional tensor transversity
amplitudes $A_{ij}$. Here the labels $i$ and $j$ denote the
transversity state $t,\perp,\parallel,0$ of the polarization vectors which
comprise the rank-two polarization tensor that was used in the computation.
We obtain for pairs $(\parallel\perp,\, t0)$, $(0\!\perp,\, t\!\perp)$,
$(0\!\parallel,\, t\!\parallel)$ the total angular momenta $J=0,1,2$,
respectively.  For the definition of the transversity amplitudes and their
general results, see \refapp{calculation} and
\refeqs{tensorAmpFirst}{tensorAmpLast}, respectively. 

At low recoil, after
application of the improved Isgur-Wise relations, we obtain
\begin{equation}
  \label{eq:}
\begin{aligned}
  A_{\parallel\!\perp,\, t0}
  & = \pm \wilson[]{T,\, T5}\, \frac{2 \kappa}{\sqrt{q^2}} M_B (1 + \hat\Lambda_0) f_0\,,
\\
  A_{t\perp,\, 0\perp}
  & = \pm \wilson[]{T,\, T5}\, \frac{\sqrt{2} \kappa}{\sqrt{q^2}} M_B (1 + \hat\Lambda_\perp)\, f_\perp\,,
\\
  A_{0\parallel,\, t\parallel}
  & = \pm \wilson[]{T,\,T5}\, \frac{\sqrt{2} \kappa}{\sqrt{q^2}} M_B (1 + \hat\Lambda_\parallel)\, f_\parallel\,,
\end{aligned}
\end{equation}
where $\hat\Lambda_{0,\perp,\parallel} = \order{\Lambda_{\rm QCD}/M_B}$
and the upper and lower sign refers to $C_T$ and $C_{T5}$, 
respectively.\\

In the presence of tensor operators T and T5 in addition to (SM + SM') the
angular observables $J_i$ receive {\it i)}~contributions
which do not interfere with other operators in $J_{1s,1c,2s, 2c, 3, 4}$,
and {\it ii)} helicity-suppressed interference contributions in $J_{1s, 1c, 5, 6s, 6c, 7}$,
and {\it iii)} no contributions in $J_{8,9}$
from the additional six transversity amplitudes $A_{t0,\, \parallel\perp,\,
  t\perp,\, t\parallel,\, 0\perp,\, 0\parallel}$.  We find
\begin{align}
    \nonumber
  \frac{8}{3}J_{1s}
  & = \Big[3 \rho_1^+ +  \rho_1^T \, (1 + \hat\Lambda_\perp)^2 \Big] f_\perp^2
\\
    \nonumber
  & + \Big[3 \rho_1^- +  \rho_1^T \, (1 + \hat\Lambda_\parallel)^2 \Big] f_\parallel^2
  + \order{\frac{m_\ell}{\sqrt{q^2}}}\,,
\\
    \nonumber
  \frac{4}{3} J_{1c}
  & = 2 \left[\rho_1^- + \rho_1^T \, (1 + \hat\Lambda_0)^2 \right]f_0^2
    + \order{\frac{m_\ell}{\sqrt{q^2}}}\,,
\\
    \nonumber
  \frac{4}{3\beta_\ell^2}(2J_{2s} & \pm J_3)
    = 2 \Big[\rho_1^\pm - \rho^T_1 \, (1 + \hat\Lambda_{\perp,\parallel})^2\Big] f_{\perp,\,\parallel}^2\,,
\\
    \nonumber
 -\frac{4}{3\beta_\ell^2}J_{2c}
     & = 2 \Big[\rho_1^- - \rho^T_1 \,(1 + \hat\Lambda_0)^2\Big] f_0^2\,,
\\
  \frac{4\sqrt{2}}{3\beta_\ell^2} J_4
     & = 2 \Big[\rho_1^- - \rho^T_1 \,(1 + \hat\Lambda_0)(1 + \hat\Lambda_\parallel)\Big] f_0 f_\parallel\,,
\\
    \nonumber
  \frac{2\sqrt{2}}{3\beta_\ell}J_5
    & = 4\, \mbox{Re} \left(\rho_2\right)f_0 f_\perp + \mathcal{O}\(\frac{m_\ell}{\sqrt{q^2}}\)\,,
\\
    \nonumber
  \frac{2}{3\beta_\ell}J_{6s}
    & = 4\, \mbox{Re} \left(\rho_2\right) f_\parallel f_\perp + \mathcal{O}\(\frac{m_\ell}{\sqrt{q^2}}\)\,,
\\
    \nonumber
  J_{6c,7}
    & = \mathcal{O}\(\frac{m_\ell}{\sqrt{q^2}}\)\,,
\\
    \nonumber
  \frac{4\sqrt{2}}{3\beta_\ell^2}J_8
  & = 4\, \mbox{Im} \left(\rho_2\right) f_0 f_\perp\,,
\\
    \nonumber
 -\frac{4}{3\beta_\ell^2}J_9
 & = 4\, \mbox{Im} \left(\rho_2\right) f_\parallel f_\perp\,.
\end{align}
Here the additional short-distance combination
reads
\begin{equation}
\begin{aligned}
  \rho^T_1 & \equiv 
    16\,\kappa^2\frac{M_B^2}{q^2} \Big(|\wilson{T}|^2 + |\wilson{T5}|^2 \Big)\,.
\end{aligned} \label{eq:rhoT}
\end{equation}
Without tensor operators the ratio $H_T^{(1)}$ is free of short- and
long-distance contributions. In the presence of the tensor operators we obtain
\begin{equation}
\begin{aligned}
  \label{eq:HT1:scen:tensor}
  H_T^{(1)}
  & = {\rm sgn}(f_0) \, {\rm sgn}(\rho_1^{-} - \rho^T_1)
\\ & \times \[1
  +  \frac{\rho_1^-\, \rho^T_1}{{2}(\rho_1^-  - \rho^T_1)^2}\, {\big(\hat\Lambda_0 - \hat\Lambda_\parallel\big)^2}\] + {\order{\hat\Lambda_i^3}}
\end{aligned}
\end{equation}
and form factor factors still cancel. Deviations from $|H_T^{(1)}| = 1$ arise at
${{\cal O} (\hat\Lambda_i^2)}$, while in $H_T^{(1b,\,1c)}$ the suppression is only
linear in ${\hat\Lambda_i}$. For instance,
\begin{align}
  H_T^{(1b)}
  & = 1 - \frac{\rho^T_1}{\rho_1^-  - \rho^T_1} {(\hat\Lambda_0 - \hat\Lambda_\parallel)}
       + \order{\hat\Lambda_i^2} \,.
\end{align}

We further find that in scenarios with tensor operators $H_T^{(3,4,5)}$ remain
free of hadronic form factors, and the relations $H_T^{(2)}/H_T^{(3)} = 1$ and
$H_T^{(4)}/H_T^{(5)} = 1$ hold up to helicity-suppressed and power-suppressed
terms, respectively, see Table \ref{tab:overview} and
\ref{tab:overviewHTi}.

The relation $J_{1c} + J_{2c} = 0$, valid in the (SM+SM') for $m_\ell\to 0$,
is broken by  $\rho_1^T$
\begin{align}
  J_{1c} + J_{2c} & = 
    3\, \rho_1^T \, (1 + {\hat\Lambda_0})^2 f_0^2 
    + \order{\frac{m_\ell}{\sqrt{q^2}}}.
\end{align} 

%
\subsection{Interference between operator sets \label{sec:interference}}

\begin{table*}
\begin{center}
\begin{tabular}{c|cccccc}
\hline \hline
\tabvsptop\tabvspbot
  Scenario
  & $|H_T^{(1)}| = 1$
  & $H_T^{(2)}=H_T^{(3)}$
  & $H_T^{(4)}=H_T^{(5)}$
  & $J_{6c} = 0$
  & $J_7 = 0$
  & $J_{8,9} = 0$
\\
\hline
\tabvsptop\tabvspbot
SM        
  & \checkmark & \checkmark & (\checkmark) & \checkmark & \checkmark & \checkmark
\\
\tabvsptop\tabvspbot
SM $+$ (S+P)
  & \checkmark    
  & $\displaystyle \frac{m_\ell}{Q}\,\Re{\wilson{79}\Delta_S^*}$
  & (\checkmark)            
  & $\displaystyle \frac{m_\ell}{Q}\,\Re{\wilson{79}\Delta_S^*}$
  & $\displaystyle \frac{m_\ell}{Q}\,\Im{\wilson{79}\Delta_S^*}$
  & \checkmark
\\
\tabvsptop\tabvspbot
SM $+$ (T+T5)
  & $\displaystyle \frac{\Lambda^2}{Q^2} \rho^T_1$ 
  & $\displaystyle \frac{m_\ell}{Q}\,\Re{\wilson{10} \wilson[*]{T(T5)}}$
  & $\displaystyle (\checkmark)$
  & $\displaystyle \frac{m_\ell}{Q}\,\Re{\wilson{10} \wilson[*]{T}}$
  & $\displaystyle \frac{m_\ell}{Q}\,\Im{\wilson{10} \wilson[*]{T5}}$
  & $\checkmark$
\\
\tabvsptop\tabvspbot
SM $+$ SM'
  & \checkmark & \checkmark & \checkmark & \checkmark & \checkmark &
  $\Im {\rho_2}$
\\
\tabvsptop\tabvspbot
all
  & $\displaystyle \frac{\Lambda^2}{Q^2} \rho^T_1$
  & $\Re{\wilson{T(T5)}\Delta_{P(S)}^*}$
  & $\displaystyle \frac{\Lambda}{Q}\,{\rho^{T}_1}\,\Im{\rho_2}$
  & $\Re{\wilson{T(T5)} \Delta_{P(S)}^*}$
  & $\Im{\wilson{T(T5)} \Delta_{S(P)}^*}$
  & $\Im{\rho_2}$
\\ \hline \hline
\end{tabular}
\end{center}
\caption{The low recoil relations in SM-like models (first row) and 
  the leading terms that break them in SM extensions.
   A \checkmark denotes at most corrections of order $\alpha_s/m_b$ and 
  $\mathcal{C}_7/(\mathcal{C}_9 m_b)$.    A (\checkmark) reminds that up to the latter corrections
  $H_T^{(4,5)}=0$.
  Here $\Lambda = \Lambda_{\rm QCD}$, $Q={\cal{O}}(m_b, \sqrt{q^2})$
  and $\Delta_{S,P}\equiv C_{S,P}-C_{S',P'}$, for details see text.
  \label{tab:overview}}
\end{table*}

\begin{table}
\begin{center}
\begin{tabular}{c|ccccc} 
\hline \hline
\tabvsptop\tabvspbot
  Scenario
  & $H_T^{(1)}$
  & $H_T^{(2)}$
  & $H_T^{(3)}$
  & $H_T^{(4)}$
  & $H_T^{(5)}$
\\
\hline
\tabvsptop\tabvspbot
SM         
  & \checkmark & \checkmark & \checkmark & ---        & ---       \\
\tabvsptop\tabvspbot
SM $+$ (S+P)
  & \checkmark & $A_0$      & \checkmark & ---        & ---       \\
\tabvsptop\tabvspbot
SM $+$ (T+T5)
  & \checkmark & \checkmark & \checkmark & \checkmark & \checkmark \\
\tabvsptop\tabvspbot
SM $+$ SM'
  & \checkmark & \checkmark & \checkmark & \checkmark & \checkmark \\
\tabvsptop\tabvspbot
all
  & \checkmark & $A_0$      & \checkmark & \checkmark & \checkmark \\
\hline \hline
\end{tabular}
\end{center}
\caption{The low recoil 
  observables $H_T^{(i)}$, and the degree to which they remain free of hadronic
  input. A \checkmark denotes at most corrections of order $\alpha_s/m_b$ and 
  $\mathcal{C}_7/(\mathcal{C}_9m_b)$, while $A_0$ denotes breaking through terms
  involving the corresponding $B\to K^*$ form factor. Observables marked with --- 
  vanish in the considered scenario.
  \label{tab:overviewHTi}}
\end{table}

When considering the complete set of $|\Delta B|=|\Delta S|=1$ semileptonic
operators, all of the previously presented low recoil relations are broken at
some level, which can however be parametrically suppressed and small.  For
instance, the relations $H_T^{(2)}/H_T^{(3)} = 1$ and $J_7 = 0$ are broken at
leading order by the simultaneous presence of tensors and scalars only, while
$H_T^{(4)}/H_T^{(5)} = 1$ remains intact up to ${\cal{O}}(\hat
\Lambda_i)$-suppressed terms, see \reftab{overview} for an overview.

The angular observables $J_{1c}, J_5,J_{6c}$ and $J_7$ receive contributions
from (pseudo-) scalar operators involving the form factor $A_0$, see Section
\ref{sec:scalar:ops}. These terms modify the otherwise general structure
\begin{align}
  \label{eq:general}
  J_a & \sim \rho_i \,f_k^{} f_l \, .
\end{align}
Corrections to $J_{1c}$ arise from operators S and P, while $J_5$, $J_{6c}$ and
$J_7$ are modified by interferences of S and P with tensor operators.

Since the angular observables $J_{2s,2c,3,4,6s,8,9}$ obey Eq.~(\ref{eq:general})
it follows that $H_T^{(1,3,4,5)}$ remain free of hadronic inputs in the complete
operator basis.  Our findings are summarized in \reftab{overviewHTi}.

%
%
\section{$\bar{B}\to \bar{K}\ell^+\ell^-$ at low recoil \label{sec:BKll}}

The decay $\bar{B}\to \bar{K}\ell^+\ell^-$ is another accessible FCNC channel,
which depends on the Wilson coefficients in a complementary way to $\bar{B}\to
\bar{K}^*\ell^+\ell^-$. The angular distribution of $\bar{B}\to \bar{K}\ell^+\ell^-$
can be written as
\begin{align}
  \frac{d^2\Gamma^K}{dq^2 d\!\cos\thl} & = 
  a + b \, \cos\thl + c \, \cos^2\!\thl \,,
\end{align}
where the angle $\thl$ is defined as in $\bar{B}\to
\bar{K}^*\ell^+\ell^-$ decays, see Appendix \ref{sec:ang:dist}. The $q^2$-dependent
coefficients $a$, $b$ and $c$ are related to the decay rate, the lepton
forward-backward asymmetry, $A_{\rm FB}^K$, and the flat term, $F_H$, as
follows \cite{Bobeth:2007dw}
\begin{equation}
\begin{aligned}
  \frac{d\Gamma^K}{dq^2} & = 2\, (a + c/3) \,, &
\\[0.2cm]
  A_{\rm FB}^K & = \frac{b}{d\Gamma/dq^2} \,, &
  F_H & = \frac{2\,(a + c)}{d\Gamma/dq^2}\,.
\end{aligned}
\end{equation}
Here we label the  $\bar{B}\to
\bar{K} \ell^+\ell^-$decay rate and forward-backward asymmetry by a superscript
'$K$' to distinguish them from the ones in $\bar{B}\to \bar{K}^*\ell^+\ell^-$
decays.

Similar to $\bar{B}\to \bar{K}^*\ell^+\ell^-$ the low recoil OPE  and the
Isgur-Wise relations can be applied, allowing to trade the $B\to K$ tensor
form factor $f_T$ for the vector one $f_+$ \cite{Beylich:2011aq,Bobeth:2011nj}.
We obtain for the extended operator basis at low recoil to leading order
in $1/m_b$
\begin{align}
  \frac{4\,a}{\Gamma_0 \sqrt{\lambda_0}^3 f_+^{\,2}} & = 
     \rho_1^+ + \frac{f_0^{\,2}}{f_+^{\,2}} \rho^{S + P}\\
     & + \frac{m_\ell}{M_B} \rho^{T\times 79} 
     + \frac{m_\ell}{\sqrt{q^2}} \frac{f_0^{\,2}}{f_+^{\,2}} \rho^{P\times 10}\,,
     \nonumber
\\
  \frac{b}{\Gamma_0 \lambda_0 (M_B^2 - M_{K}^2) f_+ f_0} & = 
  \rho^{S\times T + P\times T5}\\
  & + \frac{m_\ell}{\sqrt{q^2}} \rho^{T5\times 10} + \frac{m_\ell}{m_b\!-\!m_s} \rho^{S\times 79}\,,
  \nonumber
\\
  \frac{4\,c}{\Gamma_0 \sqrt{\lambda_0}^3 f_+^{\,2}} & = 
    -\rho_1^+ + \rho_1^T \,,
\end{align}
  where $\lambda_0
\equiv \lambda(M_B^2, M_K^2, q^2)$ and
    \begin{align}
  \rho^{S+P} & \equiv 
  \frac{q^2(M_B^2 - M_{K}^2)^2}{(m_b - m_s)^2\lambda_0}\\
  & \times \left( |\wilson{S} + \wilson{S^\prime}|^2 + |\wilson{P} + \wilson{P^\prime}|^2 \right)\,,
  \nonumber
\\
  \rho^{P\times10} & \equiv 
  \frac{\sqrt{q^2}}{(m_b - m_s)} \frac{(M_B^2 - M_{K^*}^2)^2}{\lambda_0}\\
  & \times 4\, \mbox{Re} \left[\left(\wilson{P} + \wilson{P^\prime}\right) 
                     \left(\wilson{10} + \wilson{10^\prime}\right)^\ast \right]\,,  
  \nonumber
\\
  \rho^{T\times 79} & \equiv 
  16\, \kappa \frac{M_B^2}{q^2}\, 
  \mbox{Re} \left[ \wilson{T} \left(\wilson[eff]{79} + \wilson[eff]{7'9'}\right)^*\right] \,,
\\
  \rho^{S\times T + P\times T5} & \equiv 
  2\, \kappa \frac{M_B}{m_b - m_s}\\
  & \times \mbox{Re} \left[ \left(\wilson{S} + \wilson{S^\prime}\right) \wilson[*]{T}
                 + \left(\wilson{P} + \wilson{P^\prime}\right) \wilson[*]{T5} \right] \,,
  \nonumber
\\
  \rho^{T5\times 10} & \equiv
  4\, \kappa \frac{M_B}{\sqrt{q^2}}\,
  \mbox{Re} \left[ \left(\wilson{10} + \wilson{10^\prime}\right) \wilson[*]{T5} \right] \,,
\\
  \rho^{S\times 79} & \equiv
  \mbox{Re} \left[ \left(\wilson{S} + \wilson{S'}\right) 
                   \left(\wilson[eff]{79} + \wilson[eff]{7'9'}\right)^* \right] \,.
\end{align}
Here we have neglected terms suppressed by $m_\ell^2/M_B^2$, but kept
those proportional to $m_\ell/\sqrt{q^2}$. The form factor $f_+$ and
the scalar one $f_0$, as well as the normalization $\Gamma_0$ are defined
in \cite{Bobeth:2007dw, Bobeth:2011nj}, whereas $\rho_1^+$ and $\rho_1^T$
have been introduced in Section  \ref{sec:lowreco:pheno}. 

In the scenario (SM+SM') the differential decay rate
\begin{align}
  \frac{d \Gamma^K}{dq^2} & 
  = \Gamma_0 \frac{\sqrt{\lambda_0}^3}{3} f_+^2\, \rho_1^+ \,,
\end{align}
yields a complementary constraint on $\rho_1^+$. The CP asymmetry of the
rate, $ A_{\rm CP}$, turns out to be free of long distance uncertainties,
as $f_+$ cancels, and the CP asymmetries 
\begin{align} \label{eq:acpidentical}
  A_{\rm CP} [\bar{B}\to \bar{K} \ell^+\ell^-] & 
  = a_{\rm CP}^{(1,+)}[\bar{B}\to \bar{K}^* \ell^+\ell^-]
\end{align}
in $\bar{B}\to \bar{K} \ell^+\ell^-$ and $\bar{B}\to \bar{K}^* \ell^+\ell^-$
decays are identical at low recoil, see  \refeq{aCP1plus}.
The equality Eq.~(\ref{eq:acpidentical}) allows to measure CP violation with
the combined, larger data set. If the CP asymmetries turn out to be not equal
it would imply contributions from outside of (SM+SM'). 

Furthermore, the decay $\bar{B}\to \bar{K} \ell^+\ell^-$ provides with $F_H$ a powerful
observable, which exhibits sensitivity to (S+P) and (T+T5) operators, whereas the
(SM+SM') contributions are suppressed by $m_\ell^2/q^2$ \cite{Bobeth:2007dw},
\begin{align}
  \label{eq:FH:tensor}
  F_H & = \frac{3}{2} \cdot 
    \frac{\rho_1^T + \frac{f_0^{\,2}}{f_+^{\,2}} \rho^{S+P}}
        {\rho_1^+ + \frac{1}{2} \rho_1^T + \frac{3}{2} \frac{f_0^{\,2}}{f_+^{\,2}} \rho^{S+P}}
        + \order{m_\ell / M_B}\,.
\end{align}
While the terms of $\order{m_\ell/M_B}$ in the denominator might be safely
neglected in view of the numerically leading term $\rho_1^+$, they could become
of some relevance in the numerator of $F_H$ if  $\rho_1^T$ and/or
$\rho^{S+P}$ are small and the interference of $\wilson{T}$ and/or
$\wilson{P^{(\prime)}}$ with the large (SM+SM') contributions can overcome the
lepton mass suppression. 

The importance of the flat term in SM tests and NP searches becomes manifest from
\refeq{FH:tensor} since $F_H$  is given directly by the magnitude of scalar and
tensor  Wilson coefficients and secondly it depends only on form factor ratios
rather than the form factors themselves. For tensor contributions, this residual
dependence drops out and $F_H$ is free of hadronic uncertainties. 

On the other hand, in $A_{\rm FB}^K$  tensor and (pseudo-)scalar operators need
to be either simultaneously present or their contributions are lepton-mass
suppressed, such as the interference terms between SM(') and T,T5,S, and P.
Moreover, $A_{\rm FB}^K$ depends on the ratio of form factors $f_0/f_+$.
We obtain
\begin{equation}
\begin{aligned}
  A_{\rm FB}^K &
      = \frac{3 (M_B^2 - M_K^2)}{\sqrt{\lambda_0}} \frac{f_0}{f_+}
        \frac{\rho^{S\times T + P\times T5}}
             {\rho_1^+ + \frac{1}{2} \rho_1^T + \frac{3}{2}\frac{f_0^2}{f_+^2} \rho^{S+P}}\\
    & + \order{m_\ell / M_B}\,.\\
\end{aligned}
\end{equation}

%
%
\section{Sensitivity to New Physics \label{sec:sensitivity}}

\begin{figure}
\begin{center}
    \includegraphics[width=.48\textwidth]{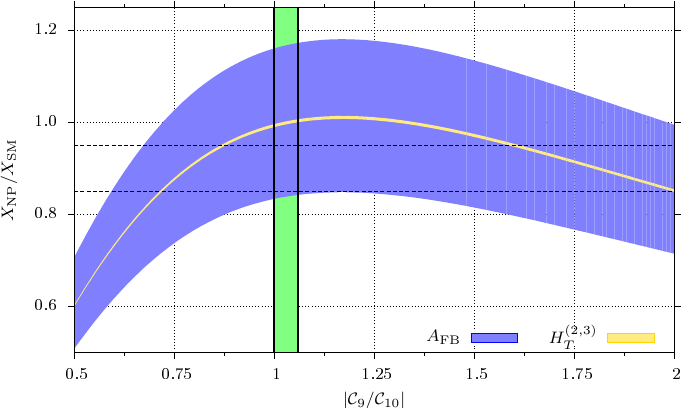}
    \caption{The sensitivity of $A_{\rm FB} $ (blue shaded band) and
      $H_T^{(2,3)}$ (thin gold shaded band) to $|\wilson{9}/\wilson{10}|$ normalized
      to their respective SM values and in the low recoil bin
      $14.18\GeV^2\!\leq\!q^2\!\leq\!19.21 \, \mbox{GeV}^2$. The dashed horizontal
      black lines indicate hypothetical measurements at $(90 \pm 5) \%$, while
      the vertical green band shows  $|\wilson{9}/\wilson{10}|$ in the SM.
        \label{fig:np-sensitivity}
    }
\end{center}
\end{figure}
\begin{figure}
\begin{center}
    \includegraphics[width=.48\textwidth]{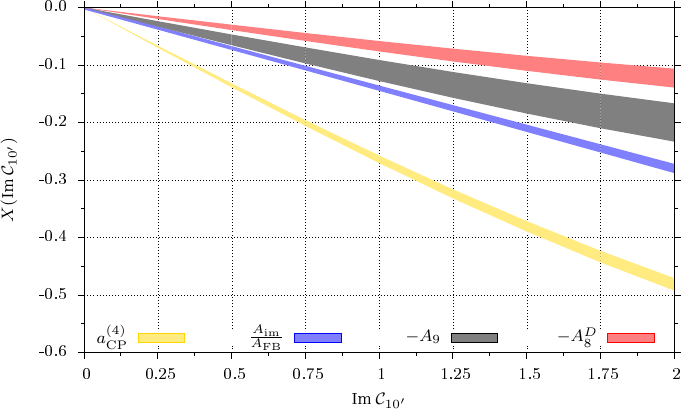}
    \caption{
        The sensitivity of  $a_{\rm CP}^{(4)}$, $A_{\rm im}/A_{\rm FB}$, $-A_9$
        and $-A_8^D$ (from bottom to top) to $\Im{\wilson{10'}}$ in the
        low recoil bin $14.18\GeV^2\!\leq\!q^2\!\leq\!19.21\GeV^2$.
        All other NP couplings including $\Re{\wilson{10'}}$ are set to zero.
        \label{fig:np-sensitivity2}
      }
\end{center}
\end{figure}

The fact that no order one NP signals have been observed in (semi)leptonic
$|\Delta B| = 1$ processes to date suggests that NP effects with the exception of null
tests are suppressed with respect to the SM contributions.
Good control of theoretical uncertainties is therefore crucial to progress.

Here we study the sensitivity of $\bar B \to \bar K^* \ell^+ \ell^-$ observables
at low recoil.  We demonstrate the advantages of the optimized low recoil
observables $H_T^{(i)}$ and related CP asymmetries.
For numerical input, see Appendix~\ref{sec:update}.

We begin studying the sensitivity of $H_T^{(2,3)}$ versus $ A_{\rm FB}$ to $\wilson{9,10}$ 
within the SM operator basis, 
where   \cite{Bobeth:2010wg}
\begin{align}
   H_T^{(2,3)} & = 2\, \frac{\rho_2}{\rho_1} \, ,\\
   A_{\rm FB} & = 3\, \frac{\rho_2}{\rho_1}
            \frac{f_\perp f_\parallel}{(f_\perp^2 + f_\parallel^2 + f_0^2)}\, ,
\end{align}
and
\begin{align}
  \frac{\rho_2}{\rho_1}
    & \sim \frac{r}{1 + |r|^2}\,, &
     & &
    r & = \frac{\wilson{9}}{\wilson{10}}\,, &
\end{align}
and $r_{\rm SM} =-1.03 \pm 0.03$.

In  \reffig{np-sensitivity} we show $H_T^{(2,3)}$ and $A_{\rm FB}$ integrated in 
the low recoil bin $14.18\GeV^2\!\leq\!q^2\!\leq\!19.21\GeV^2$ 
versus $|\wilson{9}/\wilson{10}|$ and normalized to their SM values within their
respective theory uncertainties.  
As can be seen,  a  hypothetical  measurement of $A_{\rm FB}$ as good as
$(90\pm5)\,\%$ (horizontal dashed black lines) would
not be able to distinguish the SM from NP,  due to the  theory uncertainty. The
latter is dominated by the one of the form factors, which only partially cancel
within $A_{\rm FB}$.  
Given our current understanding of the $B\to K^*$  form
factors, an experimental determination of $A_{\rm FB}$ in the low recoil region that strives to
indicate NP must exclude values larger than $0.85\times
A_{\rm FB}^{\rm SM}$ and lower than $1.18 \times A_{\rm FB}^{\rm SM}$. 
On the other hand, the  $(90\pm5)\,\%$  measurement in $H_T^{(2)}$ or 
$H_T^{(3)}$ would suffice to
establish NP due to the small, subpercent level theoretical uncertainty \cite{Bobeth:2010wg}. 
An advanced few percent-level measurement would probe 
$|\wilson{9}/\wilson{10}|$ up to a discrete ambiguity at similar, few percent level.

In   \reffig{np-sensitivity2} we show the $q^2$-integrated observables $A_{\rm im} /
A_{\rm FB}=J_9/J_{6s}$, $a_{\rm CP}^{(4)}$, $-A_8^D$ and $-A_9$ 
for $14.18\GeV^2\leq q^2\leq 19.21\GeV^2$.  
For the definition of $A_{8}^D,A_9$, see \cite{Bobeth:2008ij}. All observables are shown
as functions of the (imaginary part of the) NP coupling $\Im{\wilson{10'}}$. All other
NP Wilson coefficients including $\Re{\wilson{10'}}$ are assumed to be zero. Since the observables are odd functions
of $\Im{\wilson{10'}}$ to a very high degree  the values for
$\Im{\wilson{10'}} < 0$ are not shown.

We find that $A_{\rm im}/A_{\rm FB}$ as well as $a_{\rm
  CP}^{(4)}$ are better suited to probe small values of $\Im{\wilson{10'}}$ due
to their steeper slope and smaller relative theoretical uncertainties.
Approximately (see Eq.~(\ref{eq:aveX}) for $\braket{}$-notation),
\begin{align}
\braket{A_8^D} &\simeq (+0.061 \pm  0.008) \, \Im{\wilson{10'}} ,\\
\braket{A_9} & \simeq (+0.10 \pm 0.02) \, \Im{\wilson{10'}} ,  \\
\braket{A_{\rm im}}/\braket{A_{\rm FB}} & \simeq (-0.140 \pm 0.004) \,  \Im{\wilson{10'}} , \\ 
\braket{a_{\rm  CP}^{(4)}}  &\simeq (-0.240 \pm 0.005) \,  \Im{\wilson{10'}}  \, ,
\end{align}
where we estimated the theory uncertainty from residual $1/m_b$ corrections and form factors
similar to \cite{Bobeth:2011gi}.
Note that both $A_{\rm im}$ and $A_{\rm FB}$ have been separately measured by
CDF in two low recoil bins  \cite{Aaltonen:2011ja}. Due to the current experimental
uncertainties and the absence of information on the correlation of the individual
errors we refrain from calculating the ratio. 
 
%
\section{Probing the OPE  \label{sec:opebreak}}

The relations between the low recoil observables can be used to quantitatively
test the performance of the OPE. We employ the following ansatz:
\begin{align} \label{eq:corr-ansatz}
  A^{L,R}_i & \propto C^{L,R}\, f_{i} (1+ \epsilon_i)\, , &  i=\perp,||,0\, .
\end{align}
Here the terms $\epsilon_i$ parametrize effects beyond Eq.~(\ref{eq:universal})
to each transversity state. These include higher order power corrections or
contributions from even beyond the OPE such as duality violation.
There are no separate corrections
for the left- and right-handed lepton chiralities as the photon-current as a mediator
of the considered effects  couples vectorlike.
Although not explicitly written the
$\epsilon_i$ are in general $q^2$-dependent. 
Our ansatz parametrizes the most general situation within the SM  neglecting
lepton mass corrections of order $m_\ell^2/q^2$. 

The generic size of the subleading $1/m_b$ corrections imply $\epsilon_i$ of the
order $\alpha_s \Lambda/m_b$ or $\wilson{7}/\wilson{9}\, \Lambda/m_b$, about few
percent.  This is taken into account in current uncertainty budgets
\cite{Bobeth:2011gi}.  It is therefore desirable to have sensitivity to
corrections at the (few) percent level. On the other hand, it is hard to
quantify duality violation. While in a toy model duality violating contributions
have been estimated to be very small \cite{Beylich:2011aq} it is nevertheless
useful to have experimental checks.

We find that the corrections enter the short-distance and form factor free
relations quadratically
\begin{align} 
  \label{eq:quad}
  \left\{ |H_T^{(1)}|,\quad H_T^{(1b)},\quad \frac{H_T^{(2)}}{H_T^{(3)}} \right\}
  & = 1 + \order{\epsilon^2} ,
\end{align}
and the form factor ratios $f_i/f_k$ linearly. The null tests depend linearly on
the imaginary parts
\begin{align} 
  \label{eq:null}
  J_{7,8,9} & = \order{\mbox{Im}(\epsilon)},
\end{align}
while the real parts enter at second order only.
Note that the corrections  Eq.~(\ref{eq:quad})  vanish for
$\mbox{Im}(\epsilon_i)=0$ since for all $\epsilon_i$ real  the ansatz Eq.~(\ref{eq:corr-ansatz}) would correspond to a mere rescaling of the form factors.\footnote{We thank the unknown referee for emphasizing this point.}

The double suppression in Eq.~(\ref{eq:quad}) makes these observables sensitive
to somewhat sizable effects only or requires high experimental precision: For
$\epsilon \sim 30\,(10)\,\%$ the correction amounts to about $10\, (1)\, \%$.
On the other hand, the respective background from the SM OPE is at the permille
level.

Due to the linear dependence and the generic appearance of unsuppressed strong
phases in the nonperturbative regime we find that the null tests
Eq.~(\ref{eq:null}) have potentially higher sensitivity to OPE corrections.  Up to $
\order{\epsilon^2}$-corrections
\begin{align}
  J_7^\epsilon & = - 3\sqrt{2}\, \beta_\ell\, \rho_2\, f_0 f_\parallel \,
  \Im{\epsilon_0 - \epsilon_\parallel} \, , 
\\
    J_8^\epsilon & = \frac{3}{2 \sqrt{2} } \beta_\ell^2 \rho_1 f_0 f_\perp \Im{\epsilon_0 - \epsilon_\perp} \, ,
\\
  J_9^\epsilon & =\frac{3}{2 } \beta_\ell^2 \rho_1 f_\parallel f_\perp \Im{\epsilon_\perp - \epsilon_\parallel} \, .
\end{align}
All coefficients in front of the imaginary parts are parametrically unsuppressed
in the angular distribution.

We investigate corrections from NP in Section \ref{sec:NPpollution} and comment on an
experimental background from $K \pi$ in an S-wave with invariant mass around the
$K^*(892)$ mass in Section \ref{sec:Swave}.

%
\subsection{NP pollution \label{sec:NPpollution}}

\begin{figure}
\begin{center}
    \includegraphics[width=.40\textwidth]{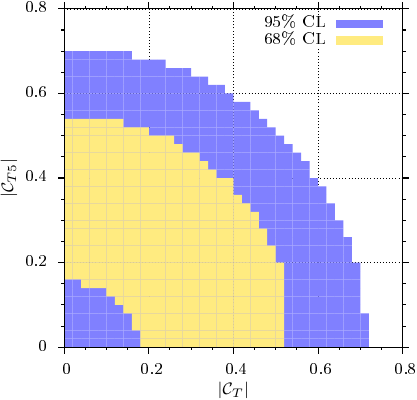}
    \caption{The constraints on $|\wilson{T}|$ and $|\wilson{T5}|$ from 
$\bar{B}\to \bar{K} \ell^+ \ell^-$ low recoil data at $68\%$~CL (inner gold area)
  and $95\%$~CL (outer blue areas). 
                  \label{fig:tensorbound}
      }
\end{center}
\end{figure}

As summarized in \reftab{overview}, beyond the SM deviations from
$|H_T^{(1)}|=1$ arise from tensor operators only. Here we estimate their contributions
allowing for complex Wilson coefficients.
The
currently strongest constraints stem from $\bar{B}\to \bar{K} \ell^+\ell^-$ decays, discussed in \refsec{BKll}.
We employ the recent LHCb measurements
\cite{Aaij:2012vr} of the branching ratio and the flat term $F_H$
Eq.~(\ref{eq:FH:tensor}), combined with branching ratio measurements from Belle
\cite{:2009zv}, BaBar \cite{:2012vwa} and CDF \cite{BKll:CDF:ICHEP:2012}.
In the scenario (T+T5+SM+SM') we find at $95\,\%$ CL 
\begin{align} 
  \label{eq;Tbound}
  |\wilson[]{T}|^2 + |\wilson[]{T5}|^2 & \lesssim 0.5 \, . 
\end{align}
The bound is
dominated by the LHCb measurement of the two highest $q^2$-bins of $F_H$.
We recall that in $F_H$ the form factor uncertainties drop out at leading order in this kinematic
regime.

The outcome of our scan leading to Eq.~(\ref{eq;Tbound})  is shown in
\reffig{tensorbound}.
Values of both $|\wilson{T}|,|\wilson{T5}|$ near zero are disfavored at  $68\,\%$ CL.
This follows from the current low recoil data on $F_H$ \cite{Aaij:2012vr}, 
which have central values at $\sim 1 \sigma$ above the SM.

Using Eqs.~(\ref{eq:rhoT}), (\ref{eq:HT1:scen:tensor}) and  (\ref{eq;Tbound})  we obtain
\begin{align} \label{eq:HT-tensors}
    \Big | |H_T^{(1)}| - 1 \Big | \lesssim 0.08\,.
\end{align}
Scalar and pseudo-scalar operators contribute predominantly constructively to
$F_H$, such that they do not invalidate an upper bound on the tensor
contributions.

The relation $H_T^{(2)}/H_T^{(3)} = 1$ receives corrections by the simultaneous
presence of tensor and scalar operators, see \reftab{overview}. The latter are
constrained by $\bar{B}_s\to \mu^+\mu^-$ decays.  Assuming that the branching
ratio $\mathcal{B}(\bar{B}_s\to \mu^+\mu^-)$ is saturated by scalar operators
${\cal{O}}_{S^{(')}}$ we find $|\Delta_S| \lesssim 0.59$ from the most recent
upper bound $\mathcal{B}(\bar{B}_s \to \mu^+\mu^-) < 4.2\times 10^{-9}$ at
$95\,\%$ CL \cite{ATLAS+CMS+LHCb:2012}.  For this bound we also consider
$B_s$--$\bar{B}_s$ mixing effects pointed out in Ref.~\cite{DeBruyn:2012wk},
allowing for ${\cal A}_{\Delta\Gamma} = -1$ and use $y_s = 0.088\pm 0.014$. 
In combination with the bounds on the tensor
couplings we obtain
\begin{equation}
    |H_T^{(2)}/H_T^{(3)} - 1| \lesssim 0.12 \, \mbox{GeV} \times N\frac{f_\parallel A_0}{f_\perp f_0} \simeq 0.06\,.
\end{equation}
Subleading effects from scalar operators alone are suppressed by $m_\ell/Q$ and
do not exceed the percent level for muons.

The relation $J_7=0$ receives corrections from NP with either
$m_\ell/Q$-suppression or require tensor contributions.  To estimate the
sensitivity to OPE corrections vs.~NP we compare the respective contributions
\begin{align}
  \frac{J_7^{\rm NP}}{J_7^\epsilon} & \simeq 
    {2} \frac{\sqrt{\lambda}}{\sqrt{q^2}}\, 
    \frac{\Im{\wilson{T} \Delta_S^* + \wilson{T5} \Delta_P^*}}
       {\rho_2 \Im{\epsilon_0 - \epsilon_\parallel}} \,
       N \frac{A_0 f_\perp}{f_0 f_\parallel} 
\\[0.2cm]
  & \lesssim\left(  \frac{0.04}{ \Im{\epsilon_0 - \epsilon_\parallel}} \right)\, .
\end{align}
We learn that $J_7$ probes the OPE as good as at the few percent level, before
tensor induced contributions can keep up. NP contributions to $J_7$ from scalar
operators interfering with the SM are suppressed by $m_\ell/Q$, which do not
exceed the percent level for muons. 
 
We further find that both $J_8$ and $J_9$ are more sensitive to NP than $J_7$,
with similar sensitivities given as
\begin{align}
  \frac{J_8^{\rm NP}}{J_8^\epsilon} & \simeq   
    \frac{ 2 \, \Im{\rho_2}}{ \rho_1 \Im{\epsilon_0 - \epsilon_\perp}} \, , 
\\[0.2cm]
  \frac{J_9^{\rm NP}}{J_9^\epsilon} & \simeq 
    \frac{-2 \, \Im{\rho_2}}{ \rho_1 \Im{\epsilon_\perp - \epsilon_\parallel}} \, ,
\end{align}
both of which are generically order one in the presence of order one CP phases,
see Eq.~(\ref{eq;imrho2}), unless the $\epsilon_i$ are ${\cal{O}}(1)$ or larger.
We conclude that $J_8$ and $J_9$ are likely to probe CP violating right-handed
currents.

In general tensor operators are absent or small in most models; they arise {\it
  e.g.,} from box-type matching conditions, can contribute to Wilson coefficients
of dipole operators via renormalization group mixing \cite{Borzumati:1999qt,
  Bobeth:2011st} or arise in models with FCNC at tree-level, such as those with leptoquarks,
  as discussed in
\cite{Bobeth:2007dw}.  In this respect the OPE-based predictions of the
short-distance independence of $H_T^{(1)}, H_T^{(2))}/H_T^{(3)}$ and $J_7$ are very clean.

%
\subsection{S-wave pollution \label{sec:Swave}}

We discuss the impact of the experimental background from $\bar K \pi$ in an
S-wave on the $\bar B \to \bar K^* (\to \bar K \pi) \ell^+ \ell^-$ angular
analysis, recently addressed in Refs.~\cite{Becirevic:2012dp, Matias:2012qz,
  Blake:2012mb} for the low $q^2$ region.

The resonant spin zero $\bar K \pi$-contributions barely affect the low recoil
region as the masses of the well established scalar kaons, notably $K_0^*(1430)$
cause their kinematical endpoint $q^2_{\max} =15 \, \mbox{GeV}^2$ to have barely
overlap with the low recoil bin $q^2 \gtrsim (14-15) \mbox{GeV}^2$
\cite{Becirevic:2012dp}. In addition there is already phase space suppression
$\propto \lambda_0^{3/2}$ below the endpoint.
A nonresonant contribution with invariant mass around $m_{K^*}$ is, however,
not suppressed by these arguments and can affect the required precision
extraction of the angular observables $J_i$ \cite{Blake:2012mb} for SM or OPE tests.

Fortunately all S-wave backgrounds can be controlled
experimentally.  The effect of an underlying spin zero component in the $\bar B
\to \bar K^* (\to \bar K \pi) \ell^+ \ell^-$ angular distribution
Eq.~(\ref{eq:anganal}) can be parametrized as follows:

\begin{itemize}
\item[\it i)] The $J_i$ for $i=3,6,9$ do not receive S-wave contributions.
\item[\it ii)] The terms $[J_i \sin 2 \theta_K ]$ for $i=4,5,7,8$ in
  Eq.~(\ref{eq:anganal}) need to be replaced by $[J_i \sin 2 \theta_K + \tilde
  J_i \sin \theta_K ]$. The $\tilde J_i$ denote interference terms; due to their
  different angular dependence they can be isolated.
\item[\it iii)] The terms $[J_{is} \sin^2\!\thK + J_{ic} \cos^2\!\thK ]$ for
  $i=1,2$ in Eq.~(\ref{eq:anganal}) need to be replaced by $[(J_{is}+ \tilde
  J_{is}) \sin^2\!\thK + (J_{ic}+ \tilde J_{ic}) \cos^2\!\thK + \tilde
  J_{isc}\cos \theta_K]$. The interference terms $\tilde J_{isc}$ can be
  identified by angular analysis. The $ \tilde J_{is}, \tilde J_{ic}$ stem from
  S-waves only and can be measured at invariant $\bar K \pi$ masses outside of
  the $K^*(892)$ peak, where the $J_{is},J_{ic}$ can be neglected.
\end{itemize}
Note that all $\tilde J_i$ depend in general on $q^2$; they incorporate resonant
and nonresonant scalar contributions.

Procedure {\it iii)} is required for all $H_T^{(k)}$ as well as observables
which are normalized to the rate. The accuracy to which the $\tilde J_{ix}$,
$i=1,2, x=s,c$ can be measured limits the experimental precision on the
$J_{ix}$.  Note that $J_{7,8,9}$ and the observables given in
Eq.~(\ref{eq:J9J6}) can be extracted without side-band measurements. $J_9$ and
the second observable in Eq.~(\ref{eq:J9J6}), which is proportional to
$J_9/J_{6s}$ do not receive contributions from S-waves at all.

%
%
\section{Conclusions \label{sec:conclusion}}

We analyze $\bar{B}\to\bar{K}^*(\to \bar{K} \pi)\ell^+\ell^-$ and $\bar{B}\to
\bar{K}\ell^+\ell^-$ decays, $\ell =e,\mu$, at low hadronic recoil in the
most general basis of semileptonic dimension-six effective couplings.  We
investigate to which extent the beneficial features obtained from angular
analysis and the OPE in the SM-like operator basis hold. We find:

The transversity observables $H_T^{(i)}$, $i=1,3,4,5$ remain in the general
case free of hadronic matrix elements and are clean tests of the SM and beyond;
for $H_T^{(2)}$ this is true if contributions from scalar operators are ignored,
see Table \ref{tab:overviewHTi}.

The form factor ratio $f_0/f_\parallel$ can be extracted by means of
Eqs.~(\ref{eq:sd-free:ffFirst}) and (\ref{eq:ff-ratio-test}), excluding 
methods based on $J_5$ if scalar operators are present. If no chirality-flipped
operators contribute the ratios Eqs.~(\ref{eq:sd-free:ffMid}) and
(\ref{eq:sd-free:ffLast}) allow for a short-distance free extraction of form
factor ratios involving $f_\perp$. There is a residual short-distance
dependence from tensor operators in Eq.~(\ref{eq:sd-free:ffFirst}), which,
however, is $\Lambda_{\rm QCD}/M_B$ suppressed.

The low recoil relations among the $H_T^{(i)}$ and $J_{7,8,9}=0$ receive
corrections from both NP, see Table \ref{tab:overview}, and contributions beyond
the leading order OPE Eq.~(\ref{eq:universal}), as given in Eq.~(\ref{eq:corr-ansatz})  and discussed in Section \ref{sec:opebreak}.
Our analysis shows that with the present $|\Delta B|=|\Delta S|=1$ constraints
$J_7$ has model-independently the highest sensitivity to the latter corrections
at the percent level, before a potential NP background kicks in. The sensitivity
in $|H_T^{(1)}| = 1$ to OPE corrections becomes comparable or better if tensor operators are
ignored. The interplay of constraints will evolve with future rare decay
measurements, and the actual sensitivity to the OPE can increase.

The observables $J_{8,9}$ are sensitive to CP violating
chirality-flipped contributions. We suggest to explore such scenarios with the
observables $H_T^{(4,5)}$ and the respective CP- and T-odd CP asymmetry $a_{\rm
  CP}^{(4)}$, all of which vanish in the SM-like basis. Further null tests are
the ratios Eq.~(\ref{eq:J9J6}). Note that one of the latter, $ J_9/J_{6s} =
A_{\rm im}/A_{\rm FB}$ has already been experimentally accessed
\cite{Aaltonen:2011ja}.

Our findings are of direct use to the high statistics studies at the LHC(b)
experiments and forthcoming high luminosity flavor factories. We look forward to
this application and future data.

\acknowledgments

DvD is grateful to Sebastian Hollenberg for comments on the manuscript, and to
the particle theory group at the Excellence Cluster ``Origin and Structure of
the Universe'' for kind hospitality during which parts of this work have been
completed.  GH gratefully acknowledges the kind hospitality of the theory group
at DESY Hamburg, where parts of this work have been done. We thank
Roman Zwicky for useful communications.

%
\appendix

%
%
\section{Angular distribution \label{sec:ang:dist}}

The differential decay rate of $\bar{B}\to\bar{K}^* (\to \bar{K}\pi)\,
\ell^+\ell^-$ can, after summing over the lepton spins, assuming an  on-shell $\bar{K}^*$
of narrow width, and integrating
over the $\bar{K}\pi$-invariant mass, be written as
\begin{equation}
\begin{split}
  \frac{8 \pi}{3} & \frac{d^4 \Gamma}{d q^2\, d\!\cos\thl\, d\!\cos\thK\, d\phi} = 
\\[0.1cm]
    & (J_{1s} + J_{2s} \cos\!2\thl + J_{6s} \cos\thl) \sin^2\!\thK
\\[0.1cm]
  + & (J_{1c} + J_{2c} \cos\!2\thl + J_{6c} \cos\thl) \cos^2\!\thK  
\\[0.2cm]
  + & (J_3 \cos 2\phi + J_9 \sin 2\phi) \sin^2\!\thK \sin^2\!\thl
\\[0.2cm] 
  + & (J_4 \cos\phi + J_8  \sin\phi) \sin 2\thK \sin 2\thl 
\\[0.2cm]
  + & (J_5 \cos\phi  + J_7 \sin\phi ) \sin 2\thK \sin\thl \, ,
\end{split}
\label{eq:anganal}
\end{equation}
with twelve angular coefficients $J_i=J_i(q^2)$ times the angular
dependence. The angles are defined as $i)$ the angle $\thl$ between $\ell^-$ and
$\bar{B}$ in the $(\ell^+\ell^-)$ center of mass system (cms), $ii)$ the angle
$\thK$ between $K^-$ and the negative direction of flight of the $\bar{B}$ 
in the $(K^-\pi^+)$ cms \footnote{%
   This corrects the  {\it description} of $\thK$ as in v3 and earlier versions of this work, as well as
   Refs.~\cite{Bobeth:2010wg,Bobeth:2011gi,Bobeth:2008ij}. De facto, in all of these works the
   description as spelled out here has already been used for all numerical and analytical studies,
   which explicitly includes the numerical implementation within EOS \cite{EOS}.
} and $iii)$ the angle
$\phi$ between the two decay planes spanned by the 3-momenta of the
$(K^-\pi^+)$- and $(\ell^+\ell^-)$-systems, respectively \cite{Bobeth:2008ij,
  Altmannshofer:2008dz, Kruger:1999xa, Kruger:2005ep}.
  
Within the (SM+SM') operator basis holds $J_{6c} = 0$. A nonvanishing $J_{6c}$
arises only from interference between the operator sets (SM+SM') and S
\cite{Altmannshofer:2008dz}, (SM+SM') and T, and P and T \cite{Alok:2010zd}.
The explicit expressions of the $J_i$ are given in Appendix \ref{sec:ang:obs}.

We denote by 
\begin{align} \label{eq:aveX}
  \braket{J_i} = \int_{q^2_{\rm min}}^{q^2_{\rm max}} dq^2\, J_i(q^2) 
\end{align}
$q^2$-integrated angular observables $J_i$ in bins between $q^2_{\rm min}$ and $q^2_{\rm
  max}$.  For composite observables $X$ we use
  $\braket{X}=X(\braket{J_i})$. We assume in the following that an S-wave background from $\bar K \pi$
around the $K^*(892)$ mass has been removed as discussed in Section
\ref{sec:Swave}.

Starting from the $q^2$-integrated decay distribution $d^3\!\braket{\Gamma} /
d\!\cos\thl\, d\!\cos\thK d\phi$ one obtains the integrated decay rate and the
three single-angular differential distributions
\begin{align}
  \label{eq:Gint}
  \braket{\Gamma} & =
    2 \braket{J_{1s}} + \braket{J_{1c}} 
  - \frac{1}{3} \left(2 \braket{J_{2s}} + \braket{J_{2c}} \right)\,, 
\\[0.2cm]
  \label{eq:dG:dphi}
  \frac{d\braket{\Gamma}}{d\phi} & =  
      \frac{\braket{\Gamma}}{2\pi} 
    + \frac{2}{3\pi} \braket{J_3} \cos 2\phi 
    + \frac{2}{3\pi} \braket{J_9} \sin 2\phi\,,
\\[0.2cm]
  \frac{d\braket{\Gamma}}{d\!\cos\thl} & =
      \braket{J_{1s}} + \frac{\braket{J_{1c}}}{2}
    + \left(\braket{J_{6s}} + \frac{\braket{J_{6c}}}{2} \right) \cos\thl 
\nonumber \\
    & \qquad+ \left(\braket{J_{2s}} + \frac{\braket{J_{2c}}}{2} \right) \cos 2\thl \,,
  \label{eq:dG:dcosthL}
\\[0.2cm]
  \frac{d\braket{\Gamma}}{d\!\cos\thK} & = 
  \frac{3}{2} \left[
       \left(\braket{J_{1s}} - \frac{1}{3} \braket{J_{2s}} \right) \sin^2\!\thK  \right.
\nonumber \\
   & \qquad \left. + \left(\braket{J_{1c}} - \frac{1}{3} \braket{J_{2c}} \right) \cos^2\!\thK
    \right]
    \label{eq:dG:dcosthK}
\end{align}
after integration over either all or the remaining two angles, respectively. 

The lepton forward-backward asymmetry $A_{\rm FB}$ can be written as 
\begin{align} \label{eq:afb}
  \braket{A_{\rm FB}} \braket{\Gamma} & 
   = \braket{J_{6s}} + \frac{\braket{J_{6c}}}{2} \, ,
\end{align}
see Eq.~(\ref{eq:dG:dcosthL}). 
The extraction of $J_{4,5,7,8}$ has been discussed in \cite{Bobeth:2008ij}.  For
alternative methods to obtain the $J_i$, see for example \cite{Kim:2007fx,
  Altmannshofer:2008dz,Matias:2012qz}.

The longitudinal
$K^*$ polarization fraction $F_L$ can model-independently be defined  as
\begin{align}
  \label{eq:dGdcK}
  \frac{1}{\braket{\Gamma}} \frac{d\!\braket{\Gamma}}{d\!\cos\thK} & =
     \frac{3}{4} \braket{F_T} \sin^2\!\thK
   + \frac{3}{2} \braket{F_L} \cos^2\!\thK \,.
   \end{align}
{}From comparison with Eq.~(\ref{eq:dG:dcosthK}) one can read off
\begin{align}
  \label{eq:defFL}
  \braket{F_L} & = \frac{1}{\braket{\Gamma}} \left(
     \braket{J_{1c}}  - \frac{1}{3}\, \braket{J_{2c}} \right)\,, \\[0.4cm]
  \label{eq:defFT}
  \braket{F_T} & = \frac{2}{\braket{\Gamma}} \left(
     \braket{J_{1s}} - \frac{1}{3} \braket{J_{2s}} \right)\,,
\end{align}
where $F_T + F_L = 1$.

In the experimental analyses by the collaborations
Belle \cite{:2009zv}, BaBar \cite{:2012vwa}, CDF \cite{BKll:CDF:ICHEP:2012}
and LHCb \cite{ Aaij:2011aa} 
the distribution
\begin{align}
  \label{eq:dGdcL}
  \frac{1}{\braket{\Gamma}} \frac{d\!\braket{\Gamma}}{d\!\cos\thl} & = 
      \frac{3}{4} \braket{F_L} (1 - \cos^2\!\thl)  
\\ \nonumber
&   + \frac{3}{8} \braket{F_T} (1 + \cos^2\!\thl) 
   + \braket{A_{\rm FB}} \cos\thl \, &
\end{align}
is at least partially  employed.
We stress that the latter is based on 
[\cf Eqs.~(\ref{eq:J1s}) - (\ref{eq:J2c})]
\begin{align}
  \label{eq:J1:mlzero}
  J_{1s} & = 3\, J_{2s} \, , & 
  J_{1c} & = -J_{2c} \, ,
\end{align}
which is broken by $m_\ell \neq 0$ and/or in the presence of S, P, T or T5
contributions. Therefore, results for $F_L$  based on Eq.~(\ref{eq:dGdcL})
do not hold in full generality.

Note that in cases where Eq.~(\ref{eq:J1:mlzero}) holds, such as the SM with
lepton masses neglected, $F_L = (|A_0^L|^2 + |A_0^R|^2)/\Gamma= -
J_{2c}/\Gamma$. Furthermore, $\braket{J_{2s}} = 3/16\, \braket{\Gamma} (1 -
\braket{F_L})$ and $\braket{J_{2c}} = - 3/4\, \braket{\Gamma} \braket{F_L}$.

%
%
\section{Angular observables \label{sec:ang:obs}}

The $J_i(q^2)$ of Eq.~(\ref{eq:anganal}) can be conveniently expressed within
the (SM+SM') operator basis with the help of seven transversity amplitudes,
$A_{0,\perp,\parallel}^{L,R}$ and $A_t$, \cite{Kruger:2005ep}. The operators S
require an additional amplitude $A_S$, whereas the set P can be absorbed into
the amplitude $A_t$ \cite{Altmannshofer:2008dz}. In the presence of tensor
operators T and T5, six additional transversity amplitudes $A_{ij}$ need to be
introduced, with $ij = \{\parallel\perp,\, t0,\, t\!\perp,\, t\!\parallel,\,
0\!\perp,\, 0\!\parallel \}$, see Appendix \ref{sec:calculation}. In the
complete basis (SM+SM') $+$ (S+P) $+$ (T+T5) we obtain
\begin{widetext}
\begin{align}
  \label{eq:J1s}
  \frac{4}{3} J_{1s} & = 
    \frac{(2 + \beta_\ell^2)}{4} \left[|\apeL|^2 + |\apaL|^2 + (L\to R) \right]
    + \frac{4 m_\ell^2}{q^2} \re\left(\apeL^{}\apeR^* + \apaL^{}\apaR^*\right)
\\ \nonumber & \hspace{0.5cm}
  + 4\, \beta_\ell^2 \big(|A_{0\perp}|^2 + |A_{0\parallel}|^2\big)
       + 4\, (4- 3 \beta_\ell^2)\, \big(|A_{t\perp}|^2 + |A_{t\parallel}|^2\big)
\\ \nonumber & \hspace{0.5cm}
  + 8\sqrt{2} \frac{m_\ell}{\sqrt{q^2}} \re\left[
      (\apaL + \apaR) A_{t\parallel}^* {+ (\apeL + \apeR) A_{t\perp}^*}
    \right] , 
\\[0.2cm]
  \frac{4}{3} J_{1c} & = 
    |\azeL|^2 +|\azeR|^2
  + \frac{4 m_\ell^2}{q^2} \Big[|A_t|^2 + 2\,\re(\azeL^{}\azeR^*) \Big]
  + \beta_\ell^2 |A_S|^2 
\\ \nonumber & \hspace{0.5cm}
  + 8\, (2 - \beta_\ell^2) |A_{t0}|^2 + 8\, \beta_\ell^2 |A_{\parallel\perp}|^2
  + 16\, \frac{m_\ell}{\sqrt{q^2}} \re\Big[ (\azeL + \azeR) A_{t0}^*\Big],
\\[0.2cm]
  \frac{4}{3} J_{2s} & =
    \frac{\beta_\ell^2}{4} \bigg[ |\apeL|^2+ |\apaL|^2 + (L\to R)
       - 16 \, \big(|A_{t\perp}|^2 + |A_{t\parallel}|^2 + |A_{0\perp}|^2 + |A_{0\parallel}|^2 \big) \bigg],
\\[0.2cm]
  \label{eq:J2c}
  \frac{4}{3} J_{2c} & =
    -\beta_\ell^2 \bigg[|\azeL|^2 + |\azeR|^2
        - 8 \, \big(|A_{t0}|^2 + |A_{\parallel\perp}|^2 \big)\bigg],
\\[0.2cm]
  \frac{4}{3} J_3 & =
    \frac{\beta_\ell^2}{2} \bigg[ |\apeL|^2 - |\apaL|^2  + (L\to R)
      + 16\, \big( |A_{t\parallel}^{}|^2 - |A_{t\perp}^{}|^2 +
      |A_{0\parallel}^{}|^2 - |A_{0\perp}^{}|^2 \big) \bigg],
\\[0.2cm]
  \frac{4}{3} J_4 & =
    \frac{\beta_\ell^2}{\sqrt{2}} \re \bigg[\azeL^{}\apaL^* + (L\to R) 
    - 8 \, \sqrt{2}\, \Big(A_{t0}^{} A_{t\parallel}^*
                         + A_{\parallel\perp}^{} A_{0\parallel}^* \Big) \bigg],
\\[0.2cm]
  \frac{4}{3} J_5 & =
    \sqrt{2}\beta_\ell\, \re \bigg[\azeL^{}\apeL^* - (L\to R)
      - 2\, \sqrt{2} A_{t\parallel}^{}  A_S^*
      - \frac{m_\ell}{\sqrt{q^2}} \Big( [\apaL + \apaR] A_S^* 
\\ \nonumber & \hspace{3.7cm}
      + 4\, \sqrt{2}\, A_{0\parallel}^{} A_t^{*} 
      {
      - 4\, \sqrt{2}\, [\azeL - \azeR]^{} A_{t\perp}^*}
      - 4\, [\apeL - \apeR] A_{t0}^* \Big) \bigg],
\\[0.2cm]
  \frac{4}{3} J_{6s} & =
    2\,\beta_\ell\, \re \bigg[\apaL^{}\apeL^* - (L\to R)
    {
    + 4\, \sqrt{2}\, \frac{ m_\ell}{\sqrt{q^2}} \Big(
[\apeL - \apeR] A_{t\parallel}^* + [\apaL - \apaR] A_{t\perp}^*}
   \Big) \bigg],
\\[0.2cm]
  \label{eq:j6c}
  \frac{4}{3} J_{6c} & =
    4\, \beta_\ell\, \re \bigg[ 2\, A_{t0}^{} A_S^* +
      \frac{m_\ell}{\sqrt{q^2}} \big[(\azeL + \azeR) A_S^* 
           + 4\, A_{\parallel\perp}^{} A_t^* \big] \bigg],
\\[0.2cm]
  \frac{4}{3} J_7 & = \label{eq:J7}
    \sqrt{2} \beta_\ell\, \im \bigg[ \azeL^{}\apaL^* - (L\to R)
      {
      + 2\, \sqrt{2} A_{t\perp}^{}  A_S^*}
      + \frac{m_\ell}{\sqrt{q^2}} \Big( [\apeL + \apeR] A_S^* 
\\ \nonumber & \hspace{3.7cm}
      + 4\, \sqrt{2}\, A_{0\perp}^{} A_t^{*} 
      + 4\, \sqrt{2}\, [\azeL - \azeR]^{} A_{t\parallel}^{*}
      - 4\, [\apaL - \apaR] A_{t0}^* \Big)\bigg],
\\[0.2cm]
  \frac{4}{3} J_8 & = 
    \frac{\beta_\ell^2}{\sqrt{2}}\, \im \bigg[
     \azeL^{}\apeL^* + (L\to R) \big) \bigg],
\\[0.2cm]
  \frac{4}{3} J_9 & =
    \beta_\ell^2\, \im \bigg[\apeL \apaL^{*} + (L\to R) \big) \bigg]\,,
\end{align}
\end{widetext}
where the lepton mass $m_\ell$ has been kept and $\beta_\ell = \sqrt{1 - 4\,
  m_\ell^2/q^2}$.

Here the transversity amplitudes contain the contributions from the operators
in Eqs.~(\ref{eq:SM:ops}) -- (\ref{eq:tensor:ops}) which are factorizable.
Non-factorizable contributions from $\Op_{i\leq 6,8}$  are taken into account by using
effective Wilson coefficients $\wilson[eff]{i}$. Within naive factorization
the transversity  amplitudes read
\begin{widetext}
\begin{align}
  A_\perp^{L,R} & = \sqrt{2} N \sqrt{\lambda}
    \Big\{\[\(\wilson{9} + \wilson[]{9'}\) \mp \(\wilson{10} + \wilson[]{10'}\)\]\frac{V}{M_B + M_{K^*}}
      + \frac{2 m_b}{q^2}\(\wilson{7} + \wilson[]{7'}\)T_1\Big\},
\\[0.2cm]
  A_\parallel^{L,R} & = -N \sqrt{2}(M_B^2 - M_{K^*}^2) \times \Big\{
\\ 
\nonumber & \hskip 2.5cm  
    \[\(\wilson{9} - \wilson[]{9'}\) \mp \(\wilson{10} - \wilson[]{10'}\)\]\frac{A_1}{M_B - M_{K^*}}
      + \frac{2 m_b}{q^2}\(\wilson{7} - \wilson[]{7'}\)T_2 \Big\},
\\[0.2cm]
  A_0^{L,R} & = -\frac{N}{2 M_{K^*}\sqrt{q^2}} \times \Big\{
\\
\nonumber & \hskip 0.5cm
    \Big[ \left(\wilson{9} - \wilson[]{9'}\right) \mp \left(\wilson{10} - \wilson[]{10'}\right) \Big]
    \Big[(M_B^2 - M_{K^*}^2 - q^2)(M_B + M_{K^*}) A_1 - \frac{\lambda}{M_B + M_{K^*}} A_2 \Big]
\\ \nonumber
    & \hskip 0.8cm + 2 m_b \left(\wilson{7} - \wilson[]{7'} \right)
        \Big[\left(M_B^2 + 3 M_{K^*}^2 - q^2\right)T_2 - \frac{\lambda}{M_B^2 - M_{K^*}^2}T_3\Big]
      \Big\}  ,
\\
  \label{eq:trAmp:At}
  A_t & = N \frac{\sqrt{\lambda}}{\sqrt{q^2}}
    \left[ 2\, (\wilson[]{10} - \wilson[]{10'}) + 
     \frac{q^2}{m_\ell} \frac{(\wilson[]{P} - \wilson[]{P'})}{(m_b + m_s)}\right] A_0,
\end{align}
\end{widetext}
\begin{align}
  \label{eq:trAmp:As}
  A_S & = -2 N \sqrt{\lambda}\, \frac{(\wilson[]{S} - \wilson[]{S'})}{(m_b + m_s)} A_0, 
\\
  \label{eq:tensorAmpFirst}
  A_{\parallel\!\perp\,(t0)} & = \pm N \frac{\wilson[]{T(5)}}{M_{K^*}}\, \Big[ 
    (M_B^2 + 3\, M_{K^*}^2 - q^2)\, T_2\\
    & \quad - \frac{\lambda}{M_B^2 - M_{K^*}^2}\, T_3
     \Big], \nonumber
\\[0.2cm]
  A_{t\perp\,(0\perp)} & 
    = \pm 2 N \, \frac{\sqrt{\lambda}}{\sqrt{q^2}}\,\wilson[]{T(5)}\, T_1, &
\\[0.2cm]
  \label{eq:tensorAmpLast}
  A_{0\parallel\,(t\parallel)} &
    = \pm 2 N \, \frac{(M_B^2 - M_{K^*}^2)}{\sqrt{q^2}}\, \wilson[]{T(5)}\, T_2. &
\end{align}
The upper and lower sign in \refeqs{tensorAmpFirst}{tensorAmpLast} refers 
$C_T$ and $C_{T5}$, respectively. The normalization factor $N$ is given as
\begin{align}
  \label{eq:trAmp:norm:factor}
  N & = G_F\, \alpha_e\, V_{tb}^{}V_{ts}^{*}\,
    \sqrt{\frac{q^2 \, \beta_\ell \,\sqrt{\lambda}}{3 \cdot 2^{10}\, \pi^5\, M_B^3}}
\end{align}
and the $B \to K^*$ form factors $V$, $A_{0,1,2}$, $T_{1,2,3}$ are defined as in
\cite{Beneke:2001at, Bobeth:2010wg, Ball:2004rg, Altmannshofer:2008dz,
  Kruger:2005ep, Alok:2010zd}.

The (SM+SM') calculation of the 4-fold differential decay rate by Kr\"uger and
Matias \cite{Kruger:2005ep} already includes the chirality-flipped operators of
the SM' basis for $m_\ell \neq 0$. We reproduce their results. The complete set
of operators  was considered in the limit $m_\ell = 0$ by Kim and Yoshikawa~\cite{Kim:2007fx}. 
The extension to $m_\ell \neq 0$ for (S+P) operators has been performed by
Altmannshofer {\it et al.} \cite{Altmannshofer:2008dz} within the transversity
amplitude formalism. We agree with the arXiv v5 of this work~\footnote{We thank
the authors of \cite{Altmannshofer:2008dz} for confirming  missing factors of 2
in Eqs.~(3.31) and (3.32) in earlier versions and the journal version.}.

The extension to $m_\ell \neq 0$ to include the tensor operators (T+T5) has been
performed by Alok {\it et al.} \cite{Alok:2009tz,Alok:2010zd}.
We agree with the arXiv v4 of reference \cite{Alok:2010zd}\footnote{We thank
Murugeswaran Duraisamy for confirming numerous typos in the journal version and the
arXiv versions prior to v4 of \cite{Alok:2010zd}.} for all expressions except for the sign of
the $A_{t\perp} A_S^*$ interference term in \refeq{J7}.

%
%
\section{$\bar{B}\to\bar{K}^*(\to \bar{K}\pi)\,\ell^+\ell^-$ Matrix Element 
  \label{sec:calculation}}

We present here the parametrization of the hadronic matrix element used to
calculate the decay $\bar{B}\to\bar{K}^*(\to \bar{K}\pi)\,\ell^+\ell^-$,
\begin{multline}
  \label{eq:BKpill:ME:X}
  {\cal M} =
  {\cal F} \Big( X_S \[\bar{\ell} \ell\]
    + X_P \[\bar{\ell} \gamma_5 \ell\]
    + X^{\mu}_V \[\bar{\ell} \gamma_{\mu} \ell\]\\
    + X^{\mu}_A \[\bar{\ell} \gamma_{\mu} \gamma_5 \ell\]
    + X^{\mu\nu}_T \[\bar{\ell} \sigma_{\mu\nu} \ell\]
    \Big)\,.
\end{multline}
We define
\begin{align}
  {\cal F} & = i \frac{\GF \alE}{\sqrt{2} \pi}\, V_{tb}^{} V_{ts}^{\ast}\,\,
               g_{K^*K\pi} D_V\, 2 |\vec{p}_K|\,,\\
\intertext{and use $\vec{p}_K$, the three momentum of the $\bar{K}$ in the $\bar{K}\pi$ cms,
}
  |\vec{p}_K| & = \frac{\sqrt{\lambda\left(M^2_{K^*}, M^2_K, M^2_\pi\right)}}{2\, M_{K^*}}\,,
\intertext{and the kinematical function $\lambda$ defined as usual}
  \label{eq:defLambda}
  \lambda(a, b, c)
    & = a^2 + b^2 + c^2 - 2(ab + ac + bc)\,.
\end{align}

\begin{table}
\resizebox{.48\textwidth}{!}{
\begin{tabular}{|llr|llr|}
\hline \hline
\tabvsptop
$A$                      & $0.812^{+0.013}_{-0.027}$       & \cite{Charles:2004jd}   &
$\lambda$                & $0.22543\pm0.00077$             & \cite{Charles:2004jd}   \\
$\bar{\rho}$             & $0.144\pm{0.025}$               & \cite{Charles:2004jd}   &
$\bar{\eta}$             & $0.342^{+0.016}_{-0.015}$       & \cite{Charles:2004jd}   \\
$\alpha_s(M_Z)$          & $0.1176$                        &                         &
$\tau_{B^+}$             & $1.638~\pico\second$            & \cite{Nakamura:2010zzi} \\
$\alpha_e(m_b)$          & $1/133$                         &                         &
$\tau_{B^0}$             & $1.525~\pico\second$            & \cite{Nakamura:2010zzi} \\
$m_c(m_c)$               & $(1.27^{+0.07}_{-0.09})~\GeV$   & \cite{Nakamura:2010zzi}   &
$M_{B^+}$                & $5.2792~\GeV$                   & \cite{Nakamura:2010zzi} \\
$m_b(m_b)$               & $(4.19^{+0.18}_{-0.06})~\GeV$   & \cite{Nakamura:2010zzi}    &
$M_{B^0}$                & $5.2795~\GeV$                   & \cite{Nakamura:2010zzi} \\
$m_t^{\rm pole}$         & $(173.3\pm1.1)~\GeV$            & \cite{:2009ec}          &
$M_{K^+}$                & $0.494~\GeV$                    & \cite{Nakamura:2010zzi} \\
$m_e$                    & $0.511~\MeV$                    & \cite{Nakamura:2010zzi} &
$M_{K^0}$                & $0.498~\GeV$                    & \cite{Nakamura:2010zzi} \\
$m_\mu$                  & $0.106~\GeV$                    & \cite{Nakamura:2010zzi} &
$M_{K^{*+}}$             & $0.89166~\GeV$                  & \cite{Nakamura:2010zzi} \\
$M_W$                    & $(80.399\pm0.023)~\GeV$         & \cite{Nakamura:2010zzi} &
$M_{K^{*0}}$             & $0.89594~\GeV$                  & \cite{Nakamura:2010zzi} \\
$\sin^2\theta_W$         & $0.23116\pm0.00013$             & \cite{Nakamura:2010zzi} &
                         &                                 & \\
\hline \hline
\end{tabular}
}
\caption{The numerical input used in our analysis. We neglect the mass of the
   strange quark. $\tau_{B^0}$ ($\tau_{B^+}$) denotes the lifetime of the neutral 
   (charged) $B$ meson. Here, $\lambda$ denotes the CKM parameter in the
   Wolfenstein parametrization.
  \label{tab:numinput}
}
\end{table}

\begin{table}[t]
\begin{tabular}{|c|c|c|}
\hline \hline
\tabvsptop
Observable & $\bar{B}^0\to\bar{K}^0\ell^+\ell^-$ & $B^-\to K^{-}\ell^+\ell^-$\\
\hline 
\tabvsptop
$10^8\times\langle BR\rangle_{4m_{\mu}^2..2.0}$       & $6.44^{+2.07}_{-1.06}$ & $6.92^{+2.22}_{-1.13}$%
\\
\tabvsptop
$10^8\times\langle BR\rangle_{2.0..4.3}$              & $7.50^{+2.56}_{-1.25}$ & $8.08^{+2.75}_{-1.35}$%
\\
\tabvsptop
$10^7\times\langle BR\rangle_{4.3..8.68}$             & $1.38^{+0.51}_{-0.25}$ & $1.48^{+0.55}_{-0.27}$%
\\
\tabvsptop
$10^7\times\langle BR\rangle_{1.0..6.0}$              & $1.63^{+0.56}_{-0.27}$ & $1.75^{+0.60}_{-0.29}$%
\\
\tabvsptop
$10^8\times\langle BR\rangle_{14.18..16.0}$           & $3.40^{+1.79}_{-0.83}$ & $3.65^{+1.92}_{-0.89}$%
\\
\tabvsptop
$10^8\times\langle BR\rangle_{16.0..18.0}$            & $3.09^{+1.76}_{-0.81}$ & $3.31^{+1.89}_{-0.87}$%
\\
\tabvsptop
$10^8\times\langle BR\rangle_{18.0..22.0}$            & $3.18^{+2.01}_{-0.92}$ & $3.41^{+2.16}_{-0.98}$%
\tabvspbot
\\
$10^8\times\langle BR\rangle_{16.0..q_{\rm max}^2}$   & $6.34^{+3.82}_{-1.75}$ & $6.80^{+4.10}_{-1.88}$%
\tabvspbot
\\
\hline
\tabvsptop
$10^1\times\langle F_H\rangle_{4m_{\mu}^2..2.0}$      & $1.03^{+0.06}_{-0.12}$ & $1.03^{+0.06}_{-0.12}$%
\\
\tabvsptop
$10^2\times\langle F_H\rangle_{2.0..4.3}$             & $2.37^{+0.18}_{-0.33}$ & $2.37^{+0.18}_{-0.33}$%
\\
\tabvsptop
$10^2\times\langle F_H\rangle_{4.3..8.68}$            & $1.24^{+0.12}_{-0.20}$ & $1.24^{+0.12}_{-0.20}$%
\\
\tabvsptop
$10^2\times\langle F_H\rangle_{1.0..6.0}$             & $2.54^{+0.20}_{-0.36}$ & $2.55^{+0.20}_{-0.36}$%
\\
\tabvsptop
$10^3\times\langle F_H\rangle_{14.18..16.0}$          & $7.04^{+1.47}_{-1.96}$ & $7.04^{+1.48}_{-1.97}$%
\\
\tabvsptop
$10^3\times\langle F_H\rangle_{16.0..18.0}$           & $6.93^{+1.66}_{-2.09}$ & $6.93^{+1.66}_{-2.09}$%
\\
\tabvsptop
$10^3\times\langle F_H\rangle_{18.0..22.0}$           & $8.17^{+2.43}_{-2.84}$ & $8.18^{+2.43}_{-2.84}$%
\\
\tabvsptop \tabvspbot
$10^3\times\langle F_H\rangle_{16.0..q_{\rm max}^2}$  & $7.75^{+2.10}_{-2.54}$ & $7.75^{+2.10}_{-2.55}$%
\\
\hline \hline
\end{tabular}
\caption{
  The SM predictions for $\bar{B}^0\to\bar{K}^0\ell^+\ell^-$ and 
  $B^-\to K^-\ell^+\ell^-$ decays in $q^2$ bins.  For the large recoil region 
  $q^2 \leq 8.68 \, \mbox{GeV}^2$, we use the QCDF results 
  \cite{Beneke:2001at,Bobeth:2007dw}, and include all known 
  power-suppressed contributions \cite{Beneke:2004dp}.
  For the low recoil region $q^2 \geq 14.18 \, \mbox{GeV}^2$ we
  use the OPE framework \cite{Bobeth:2011nj}. In both cases we
  employ $B\to K$ form factors, or extrapolations thereof, from 
  Ref.~\cite{Khodjamirian:2010vf}.
\label{tab:btokll}}
\end{table}

\begin{table}[t]
\begin{tabular}{|c|c|c|}
\hline \hline
\tabvsptop
Observable & $\bar{B}^0\to\bar{K}^{*0}\ell^+\ell^-$ & $B^-\to K^{*-}\ell^+\ell^-$\\
\hline 
\tabvsptop
$10^7\times\langle BR\rangle_{4m_{\mu}^2..2.0}$            & $2.17^{+0.44}_{-0.40}$  & $2.21^{+0.44}_{-0.40}$\\%
\tabvsptop
$10^7\times\langle BR\rangle_{2.0..4.3}$                   & $1.05^{+0.25}_{-0.23}$  & $1.15^{+0.27}_{-0.25}$\\%
\tabvsptop
$10^7\times\langle BR\rangle_{4.3..8.68}$                  & $2.46^{+0.52}_{-0.49}$  & $2.66^{+0.56}_{-0.53}$\\%
\tabvsptop
$10^7\times\langle BR\rangle_{1.0..6.0}$                   & $2.47^{+0.55}_{-0.51}$  & $2.67^{+0.60}_{-0.56}$\\%
\tabvsptop
$10^7\times\langle BR\rangle_{14.18..16.0}$                & $1.26^{+0.40}_{-0.34}$  & $1.35^{+0.43}_{-0.37}$\\%
\tabvsptop \tabvspbot
$10^7\times\langle BR\rangle_{16.0..q^2_{\rm max}}$        & $1.47^{+0.45}_{-0.39}$  & $1.57^{+0.48}_{-0.42}$\\%
\hline 
\tabvsptop
$10^1\times\langle A_{\rm FB}\rangle_{4m_{\mu}^2..2.0}$    & $1.08^{+0.22}_{-0.23}$  & $1.08^{+0.22}_{-0.23}$\\%
\tabvsptop
$10^2\times\langle A_{\rm FB}\rangle_{2.0..4.3}$           & $8.58^{+3.46}_{-3.00}$  & $7.66^{+3.15}_{-2.75}$\\%
\tabvsptop
$10^1\times\langle A_{\rm FB}\rangle_{4.3..8.68}$          &$-1.81^{+0.45}_{-0.46}$  &$-1.81^{+0.44}_{-0.46}$\\%
\tabvsptop
$10^2\times\langle A_{\rm FB}\rangle_{1.0..6.0}$           & $4.94^{+2.81}_{-2.52}$  & $4.20^{+2.57}_{-2.33}$\\%
\tabvsptop
$10^1\times\langle A_{\rm FB}\rangle_{14.18..16.0}$        &$-4.37^{+0.69}_{-0.71}$  &$-4.37^{+0.69}_{-0.71}$\\%
\tabvsptop \tabvspbot
$10^1\times\langle A_{\rm FB}\rangle_{16.0..q^2_{\rm max}}$&$-3.80^{+0.63}_{-0.67}$  &$-3.80^{+0.63}_{-0.67}$\\%
\hline 
\tabvsptop
$10^1\times\langle F_L\rangle_{4m_{\mu}^2..2.0}$           & $3.17^{+0.75}_{-0.76}$  & $3.43^{+0.78}_{-0.78}$\\%
\tabvsptop
$10^1\times\langle F_L\rangle_{2.0..4.3}$                  & $7.88^{+0.52}_{-0.61}$  & $7.96^{+0.50}_{-0.59}$\\%
\tabvsptop
$10^1\times\langle F_L\rangle_{4.3..8.68}$                 & $6.61^{+0.69}_{-0.75}$  & $6.63^{+0.68}_{-0.74}$\\%
\tabvsptop
$10^1\times\langle F_L\rangle_{1.0..6.0}$                  & $7.35^{+0.60}_{-0.70}$  & $7.46^{+0.58}_{-0.67}$\\%
\tabvsptop
$10^1\times\langle F_L\rangle_{14.18..16.0}$               & $3.63^{+0.51}_{-0.62}$  & $3.63^{+0.51}_{-0.62}$\\%
\tabvsptop \tabvspbot
$10^1\times\langle F_L\rangle_{16.0..q^2_{\rm max}}$       & $3.38^{+0.26}_{-0.33}$  & $3.38^{+0.26}_{-0.33}$\\%
\hline
\tabvsptop
$10^1\times\langle F_T\rangle_{4m_{\mu}^2..2.0}$           & $6.83^{+0.76}_{-0.75}$  & $6.58^{+0.78}_{-0.78}$\\%
\tabvsptop
$10^1\times\langle F_T\rangle_{2.0..4.3}$                  & $2.12^{+0.61}_{-0.52}$  & $2.04^{+0.59}_{-0.50}$\\%
\tabvsptop
$10^1\times\langle F_T\rangle_{4.3..8.68}$                 & $3.39^{+0.75}_{-0.69}$  & $3.37^{+0.74}_{-0.68}$\\%
\tabvsptop
$10^1\times\langle F_T\rangle_{1.0..6.0}$                  & $2.65^{+0.70}_{-0.60}$  & $2.54^{+0.67}_{-0.58}$\\%
\tabvsptop
$10^1\times\langle F_T\rangle_{14.18..16.0}$               & $6.37^{+0.62}_{-0.51}$  & $6.37^{+0.62}_{-0.51}$\\%
\tabvsptop \tabvspbot
$10^1\times\langle F_T\rangle_{16.0..q^2_{\rm max}}$       & $6.62^{+0.33}_{-0.26}$  & $6.62^{+0.33}_{-0.26}$\\%
\hline
\tabvsptop
$10^1\times\langle A_T^{(2)}\rangle_{14.18..16.0}$         &$-3.68^{+1.96}_{-1.75}$  &$-3.69^{+1.96}_{-1.75}$\\%
\tabvsptop \tabvspbot
$10^1\times\langle A_T^{(2)}\rangle_{16.0..q^2_{\rm max}}$ &$-6.03^{+1.50}_{-1.25}$  &$-6.03^{+1.50}_{-1.25}$\\%
\hline \hline
\end{tabular}
\caption{
  The SM predictions for $\bar{B}^0\to\bar{K}^{*0}\ell^+\ell^-$ and 
  $B^-\to K^{*-}\ell^+\ell^-$ decays in $q^2$ bins. For the large recoil region
  $q^2 \leq 8.68 \, \mbox{GeV}^2$, we use the QCDF results \cite{Beneke:2001at},
  and include all known power-suppressed contributions \cite{Beneke:2004dp}.
  For the low recoil region $q^2 \geq 14.18 \, \mbox{GeV}^2$ we use the low 
  recoil OPE framework \cite{Grinstein:2004vb, Bobeth:2010wg}. In both cases we
  use the $B\to K^*$ form factors, or extrapolations thereof, from \cite{Ball:2004rg}.
  Note that  $\langle F_L \rangle +\langle F_T\rangle=1$.
\label{tab:btokstarll}}
\end{table}

Using this parametrization, we obtain the hadronic tensors
\begin{gather}
    X_S = -\frac{i}{4 N} \cos\thK\, A_S\,,\\
    X_P = +\frac{i}{2 N} \cos\thK\, \frac{m_\ell}{\sqrt{q^2}}\, A_{t}\,,
\end{gather}
\begin{align}
\label{eq:XVA}
  X^{\mu}_{V,A} & = \frac{i}{4 N} \cos\thK\, \varepsilon^\mu(0) \, (A_{0}^R \pm A_{0}^L)
\\
\nonumber
  & + \frac{i}{8 N} \sin\thK\\
\nonumber
  & \times \Big(\varepsilon^\mu(+)\, e^{+i \phi}\!\left[(A_\parallel^R + A_\perp^R) \pm (A_\parallel^L + A_\perp^L) \right]\\
\nonumber
  & +
  \varepsilon^\mu(-)\, e^{-i \phi}\!\left[(A_\parallel^R - A_\perp^R) \pm (A_\parallel^L - A_\perp^L) \right]\Big),
\end{align}
\begin{align}
    \raisetag{4.4ex}
  X^{\mu\nu}_{T} & = \frac{\cos\thK}{N}
  \(\varepsilon^\mu(t)\,\varepsilon^\nu(0) A_{t0} - \varepsilon^\mu(+)\,\varepsilon^\nu(-)A_{\parallel\perp}\)
\\
\nonumber
  & + \frac{\sin\thK}{\sqrt{2} N}\, \varepsilon^\mu(t)\\
\nonumber
  & \times \(\varepsilon^\nu(+)\, e^{ i\phi}\, [A_{t\parallel}\, {+ A_{t\perp}}] + 
    \varepsilon^\nu(-)\, e^{-i\phi}\, [A_{t\parallel} {- A_{t\perp}}]\)\\
\nonumber
  & - \frac{\sin\thK}{\sqrt{2} N}\, \varepsilon^\mu(0)\\
\nonumber
  & \times \(\varepsilon^\nu(+)\, e^{i\phi} [A_{0\perp} + A_{0\parallel}] + 
    \varepsilon^\nu(-)\, e^{-i\phi}[A_{0\perp} - A_{0\parallel}]\),
\end{align}
where the polarization vectors $\varepsilon^\mu(n)$ in the $\bar{B}$ meson rest
frame read \cite{Altmannshofer:2008dz}
\begin{equation}
\label{eq:pol-vectors:eps}
\begin{aligned}
  \varepsilon^\mu(\pm) & = \frac{1}{\sqrt{2}}(0,1,\mp i,0)\,,\\
  \varepsilon^\mu(0)   & = \frac{1}{\sqrt{q^2}}(-q_z, 0, 0, -q_0)\,,\\
  \varepsilon^\mu(t)   & = \frac{1}{\sqrt{q^2}}(q_0, 0, 0, q_z) \, .
\end{aligned}
\end{equation}
We choose the $z$-axis in this frame along the $\bar{K}^*$ direction of flight
and $q_0$ ($q_z$) denotes the timelike (spacelike) component of the four
momentum $q^\mu$. The polarization vectors fulfill the completeness relations
\begin{equation}
\begin{aligned}
    g_{nn'} & = \varepsilon^\dagger_\mu(n)\, \varepsilon^\mu(n')\,,\\
 g_{\mu\nu} & = \sum_{n,n'} \varepsilon^\dagger_\mu(n)\, \varepsilon_\nu(n')\, g_{nn'}
\end{aligned}\label{eq:heldecomp}
\end{equation}
with $g_{nn'} = \mbox{diag}(+,-,-,-)$ for $n,n' = t, \pm, 0$. We use
the relation \refeq{heldecomp}  to insert the full set of polarization vectors
$\varepsilon^\mu(n)$ between the hadronic and leptonic currents, and introduce
the helicity amplitudes $H_{a n_1 \dots n_l}$ for arbitrary Dirac
structures $\Gamma^{\mu_1\dots\mu_l}$,
\begin{gather}
    \langle \bar{K}^*(k, \eta(a)) |\bar{s}\Gamma_{\mu_1\dots\mu_l} b|\bar{B}(p)\rangle\\
    \nonumber
    = \sum_{n_i,n'_i}
          \!\langle \bar{K}^*(k,\eta(a))|\bar{s}\Gamma^{\nu_1\dots\nu_l}b|\bar{B}(p)\rangle
      \prod_{i=1}^l \varepsilon^{\dagger}_{\nu_i}\!(n_i) g^{n_i n'_i} \varepsilon_{\mu_i}\!(n'_i)\\
    \equiv \sum_{n_i}
    H^\Gamma_{a n_1\dots n_l} \prod_{i=1}^l g^{n_i n_i} \varepsilon_{\mu_i}\!(n_i)\,.
\end{gather}
The tensorial transversity amplitudes $A_{ij}$ are related to the helicity
amplitudes $H_{a n_1 n_2}$ by means of
\begin{align}
    A^\Gamma_{0\perp}     & = \frac{1}{2} \big(H^\Gamma_{+0+} + H^\Gamma_{-0-}\big) &
    A^\Gamma_{0\para}     & = \frac{1}{2} \big(H^\Gamma_{+0+} - H^\Gamma_{-0-}\big)\nn\\
    A^\Gamma_{t\perp}     & = \frac{1}{2} \big(H^\Gamma_{+t+} - H^\Gamma_{-t-}\big) &
    A^\Gamma_{t\para}     & = \frac{1}{2} \big(H^\Gamma_{+t+} + H^\Gamma_{-t-}\big)\nn\\
    A^\Gamma_{\para\perp} & = H^\Gamma_{0+-} &
    A^\Gamma_{t0}         & = H^\Gamma_{0t0}
\end{align}
and
\begin{equation}
    A_{ij} = 2\, N \sum_{\Gamma=T,T5} C_\Gamma  A^\Gamma_{ij}\,.
\end{equation}
Note that the factor 2 above emerges from the relation $H_{a ij}^\Gamma =
 - H_{a ji}^\Gamma$, which is due to the asymmetry of $\sigma^{\mu\nu}$ under
 $\mu\leftrightarrow \nu$.
The polarization vectors of the $\bar{K}^*$ for polarizations $a = \pm,0$ in the $\bar B$ cms read
\begin{equation}
\label{eq:pol-vectors:eta}
\begin{aligned}
  \eta^\mu(\pm) & = \frac{1}{\sqrt{2}}(0,1,\pm i,0)\,,\\
  \eta^\mu(0)   & = \frac{1}{M_{K^*}}(-q_z, 0, 0, M_B-q_0) \, .
\end{aligned}
\end{equation}
This approach generalizes the concept of the transversity amplitudes,
cf. e.g. Refs.  \cite{Faessler:2002ut,Kruger:2005ep,Altmannshofer:2008dz}, to
which we also refer for the definition of the remaining transversity amplitudes
$A_{i}$, $i=0,\perp,\parallel,t,S$.

We employ $\gamma_5 = i/(4\,!)\, \varepsilon_{\alpha\beta\mu\nu} \gamma^\alpha
\gamma^\beta \gamma^\mu \gamma^\nu$, such that
\begin{equation}
\label{eq:gamma5rel}
\begin{aligned}
  \mbox{Tr}[\gamma^\alpha \gamma^\beta \gamma^\mu \gamma^\nu \gamma_5] & =
     4\, i\, \varepsilon^{\alpha\beta\mu\nu}\, ,\\
  \sigma^{\alpha\beta} \gamma_5 & = 
     -\frac{i}{2}\,\varepsilon^{\alpha\beta\mu\nu} \sigma_{\mu\nu}
\end{aligned}
\end{equation}
with $\sigma_{\mu\nu} = i/2\, [\gamma_\mu,\,
\gamma_\nu]$, and $\varepsilon_{0123} = -\varepsilon^{0123} = 1$.

%
%
\section{$\bar B \to \bar K^{(*)} \ell^+ \ell^-$ SM  predictions \label{sec:update}}

We update our SM predictions for  $\bar B\to \bar K^*\ell^+\ell^-$
\cite{Bobeth:2010wg,Bobeth:2011gi} and $\bar B\to \bar K \ell^+\ell^-$ decays
\cite{Bobeth:2011nj}.  This includes the following improvements to the EOS
\cite{EOS} source code:
Firstly, a common set of numerical input parameters  is used, given in \reftab{numinput}.
The resulting changes with respect to previous works are, however, marginal.  
Secondly, we improved the implementation of the subleading corrections within
QCDF to the amplitudes in the region of large hadronic recoil. This concerns, 
in particular,  the analytic expressions for the convolution integrals that
involve the kaon light cone distribution amplitudes. The analytic results turn
out to be more numerically stable  than previous ones. In this process, we further
switched  from $\overline{\rm MS}$ to the charm pole mass.
Lastly, we  implemented all numerically relevant power-suppressed hard scattering
and weak annihilation contributions \cite{Beneke:2004dp}. Subleading 
$V_{ub}^{} V_{us}^*$ contributions are included in the numerical analysis;
Non-factorizable effects at low $q^2$ estimated in \cite{Khodjamirian:2010vf}
are not included. The results  are presented in \reftab{btokll} and \ref{tab:btokstarll}.
We recall that the $q^2$-region of validity within QCDF is approximately within
$(1 - 7) \, \mbox{GeV}^2$. Numerical predictions in the tables are extrapolations thereof and
provided to match the experimental binning.
Note that  $\langle F_L \rangle +\langle F_T\rangle=1$  [\cf Eq.~(\ref{eq:dGdcK})], however, for
convenience  we give predictions for both observables. Since we calculated $\langle F_{L,T} \rangle$ individually the sum rule served as an independent  check of the EOS code.

Besides the improvements mentioned above all details entering our low and large recoil
predictions are given in \cite{Bobeth:2010wg,Bobeth:2011gi,Bobeth:2011nj}.
Detailed SM predictions for $H_T^{(1,2,3)}$ are given in  \cite{Bobeth:2010wg,Beaujean:2012uj}.

%

\end{document}